\documentclass[onecolumn,12pt]{article}
\pdfoutput=1
\usepackage{jheppub}
\usepackage{ifpdf}
\usepackage[bb=boondox]{mathalfa}
\graphicspath{ {figures/} }
\usepackage{amsfonts}
\usepackage{amsmath}
\usepackage{amssymb}
\usepackage{epsfig}
\usepackage{tensor}
\usepackage{pdfpages}
\usepackage{bbm}
\usepackage{graphicx,epstopdf}
\usepackage[numbers]{natbib} 
\usepackage[makeroom]{cancel}
\usepackage{hyperref}
\usepackage{array}
\usepackage[export]{adjustbox}

\usepackage[normalem]{ulem}
\usepackage{romannum}

\usepackage{ulem}

\newcommand{\RN}[1]{%
	\textup{\uppercase\expandafter{\romannumeral#1}}%
}
\newcommand{\Q}{\mt{RN}}
\newcommand{\eg}{{\it e.g.,}\ }
\newcommand{\ie}{{\it i.e.,}\ }
\newcommand{\viz}{{\it viz.,}\ }
\newcommand{\mt}[1]{\textrm{\tiny #1}}
\newcommand{\reef}[1]{(\ref{#1})}

\newcommand{\GN}{G_\mt{N}}

\newcommand{\beq}{\begin{equation}}
	\newcommand{\eeq}{\end{equation}}
\newcommand{\beqa}{\begin{eqnarray}}
	\newcommand{\eeqa}{\end{eqnarray}}

\newcommand{\bea}{\begin{eqnarray}}
	\newcommand{\eea}{\end{eqnarray}}

\newcommand{\ket}[1]{\left| #1 \right>}

\renewcommand{\(}{\left(}
\renewcommand{\)}{\right)}
\renewcommand{\[}{\left[}
\renewcommand{\]}{\right]}

\newcommand{\mC}{\mathcal{C}}

\newcommand{\mR}{\mathcal{R}}

\newcommand{\mL}{\mathcal{L}}

\newcommand{\Scft}{\Sigma_\mt{CFT}}
\newcommand{\tlam}{\tilde{\lambda}}
\newcommand{\rmin}{r_{\mt{min}}}
\newcommand{\tl}{\tilde{\lambda}}
\newcommand{\veps}{\varepsilon}

\subheader{\today}
\title{Complexity$=$Anything: Singularity Probes}

\author[a,b]{Eivind J\o rstad,}
\author[a]{Robert C. Myers}
\author[c]{and Shan-Ming Ruan}

\affiliation[a]{Perimeter Institute for Theoretical Physics, \\
	Waterloo, ON N2L 2Y5, Canada}
\affiliation[b]{Department of Physics \& Astronomy, University of Waterloo, \\
	Waterloo, ON N2L 3G1, Canada}
\affiliation[c]{Center for Gravitational Physics and Quantum Information, \\
	Yukawa Institute for Theoretical Physics, Kyoto University,\\
	Kitashirakawa Oiwakecho, Sakyo-ku, Kyoto 606-8502, Japan}

\emailAdd{ejorstad@perimeterinstitute.ca}
\emailAdd{rmyers@perimeterinstitute.ca}
\emailAdd{ruan.shanming@yukawa.kyoto-u.ac.jp}

\abstract{We investigate how the complexity=anything observables proposed by \cite{Belin:2021bga,Belin:2022xmt} can be used to investigate the interior geometry of AdS black holes. In particular, we illustrate how the flexibility of the complexity=anything approach allows us to systematically probe the geometric properties of black hole singularities. We contrast our results for the AdS Schwarzschild and AdS Reissner-Nordstr\"{o}m geometries, \ie for uncharged and charged black holes, respectively. In the latter case, the holographic complexity observables can only probe the interior up to the inner horizon.}

\begin{document}

\begin{flushright}
	\hfill{ YITP-23-41}
\end{flushright}

\vskip -2.2em

\maketitle

\section{Introduction}

Recent research on the intersection of quantum information theory with quantum gravity has revealed quantum complexity as an interesting entry in the holographic dictionary, \eg see \cite{Susskind:2014moa,Susskind:2018pmk}. In terms of the boundary field theory, quantum complexity is envisaged to be a measure of how difficult it is to prepare the boundary state from a simple reference state using a set of elementary gates. In terms of the bulk gravitational theory, the discussion of quantum complexity has drawn attention to new kinds of observables which are anchored to a boundary time slice. The three proposals that have been studied most extensively are: complexity=volume (CV) \cite{Susskind:2014rva,Stanford:2014jda},  complexity=action (CA) \cite{Brown:2015bva,Brown:2015lvg} and complexity=spacetime volume (CV2.0) \cite{Couch:2016exn}. 

However, this discussion recently saw an enormous expansion with the introduction of a broad class of new gravitational observables \cite{Belin:2021bga,Belin:2022xmt}. All of these observables exhibit two universal features, which are argued to hold for any definition of quantum complexity in a holographic setting. First, for late times in the thermofield-double boundary state, the complexity grows linearly in time reflecting the growth of the wormhole for the dual two-sided AdS black hole \cite{Susskind:2014moa,Susskind:2018pmk}. This growth continues far beyond the times at which entanglement entropies have thermalized \cite{Hartman:2013qma} and the growth rate is proportional to the black hole mass.\footnote{The latter applies to planar vacuum black holes. When extra scales such as boundary curvature or a chemical potential are introduced, the growth rate being proportional to the mass only applies as the leading result for very large black holes (see, \eg \cite{Carmi:2017jqz}).}  The second feature, known as the switchback effect, is a universal time delay in the response of complexity to perturbations of the state in the far past. These perturbations are represented by the insertion of shock waves in the bulk geometry  \cite{Stanford:2014jda}. Given the breadth of the new class of observables, this new approach was (playfully) denoted as {\it complexity=anything}, which we adopt in the following. 

A primary motivation for studying quantum complexity in holographic settings is to better understand black hole interiors from the perspective of the boundary theory. Of course, one conspicuous feature of interior geometry is the inevitable formation of spacetime singularities \cite{Hawking:1973uf}. This paper takes some first steps in investigating how the complexity=anything observables introduced by \cite{Belin:2021bga,Belin:2022xmt} interact with black hole singularities, \ie provide probes of the spacetime geometry in the vicinity of the singularity. We investigate how the flexibility of the complexity=anything approach allows us to systematically probe the geometric properties of a black hole singularity. In particular, comparing the growth rates for different gravitational observables allows us to devise a picture of the interior geometry.

However, we begin with a puzzle that first appeared in \cite{Belin:2021bga}. There it was found that a particular codimension-one observable only yielded extremal surfaces at late times with a limited range of a certain higher curvature coupling. Examining this in more detail, we find that if we tune the coupling beyond the allowed range, the complexity appears to grow linearly for a long time but after this no sensible results are evident. However, a careful examination reveals that the correct extremal surfaces are pushed to the boundary of the allowed phase space, \ie they are pushed to the black hole singularity. Hence, this serves as an indication that the spacelike singularity plays an important role in determining the maximal surface for many of the new gravitational observables.

The rest of the paper is organized as follows: In section \ref{recap}, we briefly review the complexity=anything approach introduced \cite{Belin:2021bga,Belin:2022xmt}. Section \ref{sec:end} introduces and resolve the puzzle noted above which arose in the discussion of codimension-one observables in \cite{Belin:2021bga}.  In section \ref{sec:CMC}, we consider a specific example of the complexity=anything proposal, which reduces to the geometric features of constant mean curvature surfaces. We illustrate how various properties of the spacetime singularity can be systematically revealed with this class of gravitational observables. We close the paper with a discussion of the implications of our results as well as future research directions in section \ref{sec:disc}. In appendix \ref{sec:nearsingularity}, we consider the finiteness of the Gibbons-Hawking-York boundary term for the wide variety of spacelike singularities. In appendix \ref{sec:extremalK}, we investigate the extremal surfaces associated with a particular codimension-one functional constructed by the trace of extrinsic curvature. 

\section{Complexity = Anything} \label{recap}

\begin{figure}[t!]
	\centering
	\includegraphics[width=3.25in]{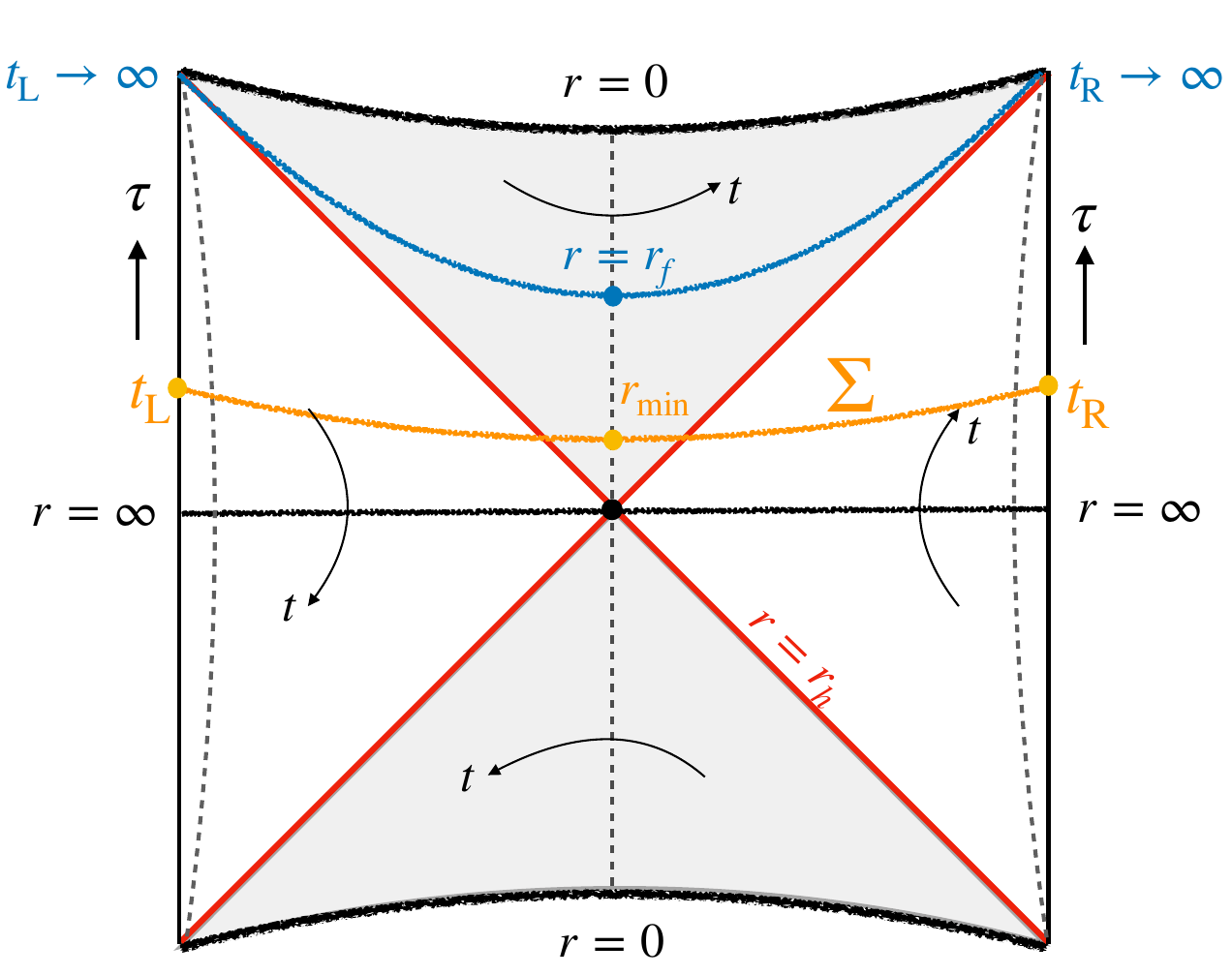}
	\includegraphics[width=2.65in]{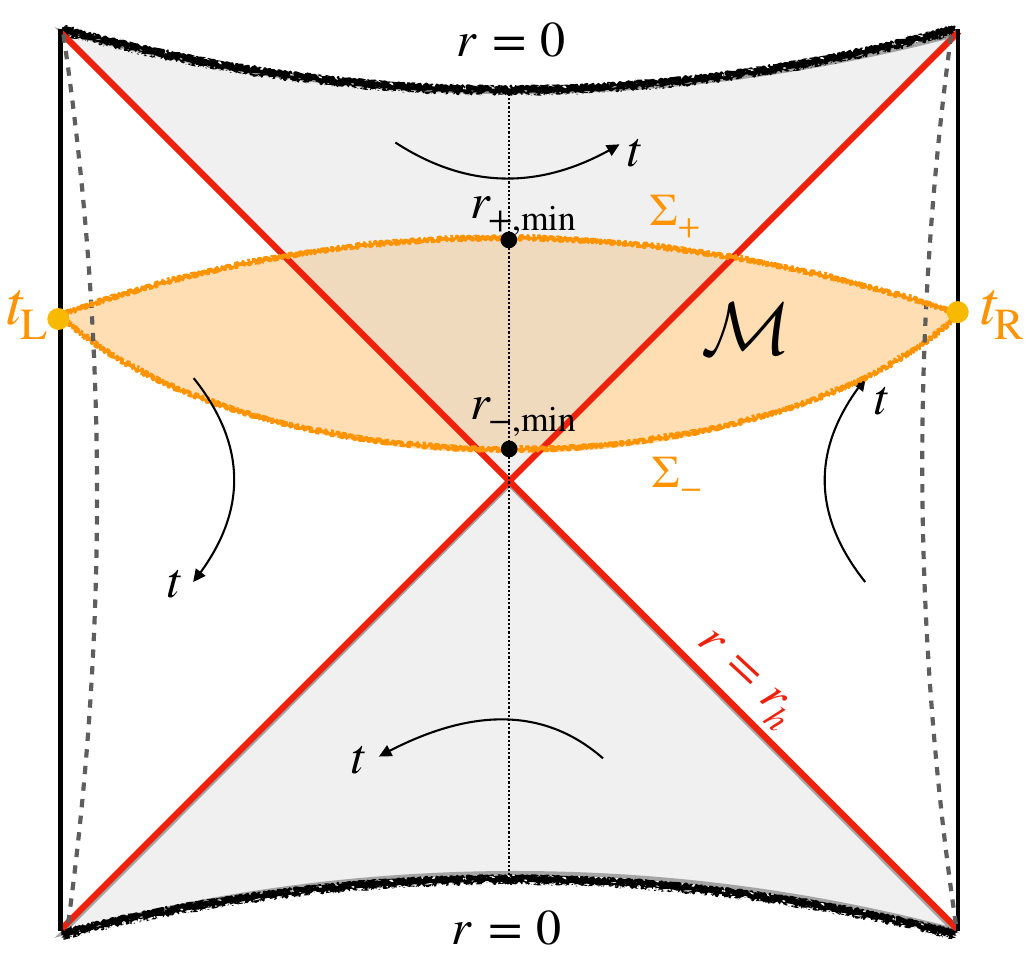}
	\caption{Left: the orange curve denotes the codimension-one extremal hypersurfaces {$\Sigma(\tau)$} associated with the complexity=anything proposal in eq.~\eqref{eq:Ccodimensionone}. Right: The orange region represents the codimension-zero subregion associated with the complexity=anything proposal.}\label{fig:TFDbalckhole}
\end{figure}

A large class of new gravitational observables was introduced in \cite{Belin:2021bga,Belin:2022xmt}, all of which appear to be equally viable candidates for holographic complexity. That is, all of these diffeomorphism-invariant observables exhibit linear late-time growth and the switchback effect in AdS black hole backgrounds. In the following, we first discuss the observables defined on codimension-one surfaces with
\begin{equation}\label{eq:Ccodimensionone}
\mC_{\rm{gen}}\(\Scft\) =\max_{\partial\Sigma=\Scft} \[ \frac{1}{G_{\mt{N}} \,L}  \int_{\Sigma} \! d^d\sigma \,\sqrt{h} \,F(g_{\mu\nu},\mathcal{R}_{\mu\nu\rho\sigma}, \nabla_\mu)  \]\,.
\end{equation}
The integral is extremized over all spacelike bulk surfaces that are asymptotically anchored to a fixed time slice $\Scft$ in the boundary theory -- see the left panel of figure \ref{fig:TFDbalckhole}. Generally, this description will be sufficient for our discussion. However, we have implicitly made two simplifications: First, the scalar function $F$ depends only on $(d+1)$-dimensional curvature invariants of the bulk geometry. However, in general, we might also include extrinsic curvatures in constructing $F$. The second specialization is that the function $F$ is used to determine the extremal surface in eq.~\reef{eq:Ccodimensionone}, and the same function appears in evaluating the observable on the extremal surface. As explained in \cite{Belin:2021bga}, two independent functions could appear in these two separate roles. We review the properties of these codimension-one observables further in section \ref{codone}.

Ref.~\cite{Belin:2022xmt} extended the complexity=anything proposal to an infinite family of gravitational observables associated with codimension-zero regions $\mathcal{M}$ anchored to a boundary time slice $\Scft$, as depicted in the right panel of figure \ref{fig:TFDbalckhole}. The bulk subregion $\mathcal{M}$ is specified by its future and past boundaries denoted $\Sigma_\pm$, \ie $\partial \mathcal{M}= \Sigma_+ \cup \Sigma_-$. The codimension-zero version of complexity=anything can then be expressed as 
\begin{equation}\label{eq:CgenZero}
	\begin{split}
		\mC_{\rm{gen}} (\Scft) &= \max_{\partial\Sigma_\pm=\Scft} \left[\frac{1}{G_N L^2 } \int_{\mathcal{M}_{G,F_{\pm}}}\!\!\!\!\!\!\! d^{d+1}x \,\sqrt{g} \ G(g_{
			\mu\nu})   \right. \\
		&\left.  +\frac{1}{\GN L }\int_{ \Sigma_+} d^d\sigma \,\sqrt{h} \,F_{+}(g_{\mu\nu}; X^{\mu}_+)  +\frac{1}{\GN L }\int_{   \Sigma_-}\! d^d\sigma \,\sqrt{h} \,F_{-}(g_{\mu\nu}; X^{\mu}_-)  \right]\,,
	\end{split}
\end{equation}
where $X^{\mu}_\pm=0$ are the embedding functions of the boundaries $\Sigma_\pm$. Hence, the maximization procedure involves varying these embeddings independently to extremize the two boundary integrals involving the scalar functionals $F_{\pm}$ as well as the bulk integral involving an independent functional $G(g_{\mu\nu})$. The observable is then given by evaluating these integrals on the extremal subregion, denoted by $\mathcal{M}_{G,F_{\pm}}$. However, we again note that the most general observables in the complexity=anything proposal would introduce an independent set of functionals to evaluated on the extremal subregion. We discuss these codimension-zero observables further in section \ref{sec:generalization}.
 
For the following discussion and the analysis in the subsequent sections, we will consider asymptotically AdS black hole backgrounds of the form
\begin{equation}\label{metricBH}
	d s^2=-f(r) d t^2+\frac{d r^2}{f(r)}+r^2 d \Omega_{k, d-1}^2\,, 
\end{equation}
where the $k \in \{0, \pm 1\}$ indicates the curvature of the $(d-1)$-dimensional line element $d \Omega_{k, d-1}^2$. For example, for $k=+1$, the spatial boundary geometry is a $(d-1)$-dimensional sphere $S^{d-1}$ of unit radius. The precise form of the blackening factor $f(r)$ will not be important for much of the discussion, but we assume that there is a horizon at $r=r_h$, \ie $f(r=r_h)=0$. The temperature of the black hole is then given by
\begin{equation}
	T_{\mathrm{BH}}=\frac{1}{4 \pi}\,\frac{d\,\! f}{d r}\bigg|_{r=r_h}\,. \label{temper}
\end{equation}
Further, for the geometry to be asymptotically AdS, we have $f(r)\simeq
\frac{r^2}{L^2}+\cdots$ as $r\to\infty$ where $L$ is the AdS curvature scale. In order to cover both the exterior and interior of the horizon, it is more convenient to work on Eddington–Finkelstein coordinates
\begin{equation}\label{eq:AdSBH}
	d s^{2}=-f(r)\, d v^{2}+2\,dv\,dr+r^{2}\, d{\Omega}^{2}_{k,d-1}\,, 
\end{equation}
where the infalling coordinate is given by $v=t + r_*(r)$ with $r_*(r)= -\int^\infty_r {d\tilde{r}}/{f(\tilde{r})}$. 

The general metric \reef{metricBH} allows us to consider quite general backgrounds, including the charged AdS Reissner-Nordstr\"{o}m black hole in section \ref{sec:CMC}. While we leave $f(r)$ general in the following, it is good to keep the vacuum AdS black hole solutions in mind as an example, with
\begin{equation}\label{blacken}
	f(r)= k+ \frac{r^2}{L^2}- \frac{\omega^{d-2}}{r^{d-2}} \,.
\end{equation}
The parameter $\omega$ determines the mass of the black hole with
\begin{equation}
	M=\frac{(d-1) \Omega_{k, d-1}}{16 \pi \GN}\, \omega^{d-2} \,,
	\label{mass}
\end{equation}
where $ \Omega_{k,d-1}$ is the {\it dimensionless} volume of the $(d-1)$-dimensional spatial boundary geometry, \eg see \cite{Chapman:2016hwi,Carmi:2017jqz}.\footnote{For example,  for a spherical boundary geometry with $k=+1$, $\Omega_{1,d-1}=2\pi^{d/2}/\Gamma(d/2)$.} 
Furthermore, this mass parameter is related to the position of the black hole horizon $r_h$ with  $\omega^{d-2}=r_h^{d-2}\( k+r_h^2/L^2 \)$. Of course, the full two-sided bulk geometry is dual to  two decoupled CFTs (on spatial geometries with constant curvature) entangled in the thermofield double (TFD) state, \ie 
\begin{equation}\label{eq:TFD}
	\left|\psi_{\mathrm{TFD}}\left(t_{\mathrm{L}}, t_{\mathrm{R}}\right)\right\rangle=\sum_{E_n} e^{-\beta E_n / 2-i E_n\left(t_{\mathrm{L}}+t_{\mathrm{R}}\right) / 2}|E_n\rangle_{\mt{L}} \otimes|E_n\rangle_{\mt{R}} \,.
\end{equation}
It is obvious that the state is invariant under the time translation $t_{\mt{R}} \to t_{\mt{R}} +\Delta t, t_{\mt{L}} \to t_{\mt{L}} -\Delta t$. Without loss of generality, we will focus on the boundary time slices at $t_{\mt{R}}=t_{\mt{L}}=\tau/2$, as illustrated in figure \ref{fig:TFDbalckhole}.

\subsection{Codimension-One Observables} \label{codone}
Strong evidence for the infinite family of codimension-one observables in eq.~\reef{eq:Ccodimensionone} can be considered as candidates for the holographic dual of complexity is that they can exhibit linear growth at late times, as expected for the time evolution of circuit complexity for the dual thermofield double state \cite{Brown:2017jil,Susskind:2018pmk,Haferkamp:2021uxo}. The details for this derivation have been presented in \cite{Belin:2021bga,Belin:2022xmt}. Here we briefly review the previous results for later reference. The goal is to show the linear growth of these infinite observables at late times, \ie 
\begin{equation}
 \lim\limits_{\tau\to\infty} 	\mathcal{C}_{\rm gen}(\tau) \sim P_{\infty}\, \tau  \,,
\end{equation}
with the growth rate $P_{\infty}$ given by a finite constant. Instead of dealing with the generalized volume $\mC_{\rm{gen}}$ which needs UV regularization, we can prove the linear growth by taking its time derivative and showing that this rate approaches a constant at late times, namely
\begin{equation}
	\lim_{\tau \to \infty}  \(    \frac{d	\mC_{\rm gen}}{d\tau} \)  = \text{constant} \,.
\end{equation}

Let us start with the simplest case defined in eq.~\eqref{eq:Ccodimensionone}. Thanks to the symmetries of the black hole geometry \reef{metricBH}, we can introduce one parameter $\sigma $ as the radial coordinate on the worldvolume of $\Sigma$ and parametrize the spacelike hypersurfaces $\Sigma$ in terms of $(v(\sigma), r(\sigma),\vec{\Omega}_{k, d-1})$. That is, the surfaces fill the transverse spatial directions respecting the symmetry of the boundary geometry but they have a nontrivial profile in the $v$ and $r$ directions. Now, the codimension-one observables \reef{eq:Ccodimensionone} can be recast as 
\begin{equation}\label{eq:defineV}
\mC_{\rm{gen}}=\frac{\Omega_{k, d-1} L^{d-2} }{\GN} \int_\Sigma d\sigma\,\(\frac{r}{L}\)^{d-1}\!\sqrt{-f(r){\dot v}^2+2\dot v\,\dot r}\ a(r)\,,
\end{equation}
where the dots denote derivatives with respect to $\sigma$. Here the factor $a(r)$  corresponds to the scalar function $F(g_{\mu\nu},\mathcal{R}_{\mu\nu\rho\sigma}, \nabla_\mu)$, which is only a function of the radius $r$ because of the symmetries of the background geometry in eq.~\reef{eq:AdSBH}. 

Finding the extremal surfaces with respect to the functional $\mC_{\rm{gen}}$ is then equivalent to solving the classical equations of motion with a Lagrangian $\mL_{\rm{gen}} \propto \mC_{\rm{gen}}$.  The conserved momentum $P_v(\tau)$ (conjugate to the infalling coordinate $v$) is given by\footnote{We have dropped the prefactor $\Omega_{k, d-1} L^{d-2} /\GN$ for convenience in defining $\mL_{\rm{gen}}$.}
\begin{equation}\label{eq:vmoment}
	P_v=\frac{\delta \mL_{\rm{gen}}}{\delta\,\dot v}
	=\frac{a(r)\,(r/L)^{d-1}\left(\dot r -f(r)\,\dot v\right)}{\sqrt{-f(r){\dot v}^2+2\dot v\,\dot r}}=\dot r -f(r)\,\dot v\,,
\end{equation}
where the second equality follows from our gauge-fixing condition, \viz 
\begin{equation}\label{eq:gauge01}
	\sqrt{-f(r){\dot v}^2+2\dot v\,\dot r} = 
	a(r)\(\frac{r}{L}\)^{d-1}\,.
\end{equation}
The extremization equation of the extremal surface $\Sigma$ then reduces to the classical equation of a non-relativistic particle \cite{Belin:2021bga}, \ie 
\begin{equation}\label{eq:Hamiltonian}
\dot{r}^2 +U_0(r) = P_v^2  \,,
\end{equation}
where the effective potential $U_0$ is defined as
\begin{equation}
	U_0(r)= -f(r)\, a^2(r) \(\frac{r}{L}\)^{2(d-1)} \,. \label{potent}
\end{equation}

\begin{figure}[t!]
	\centering
	\includegraphics[width=3in]{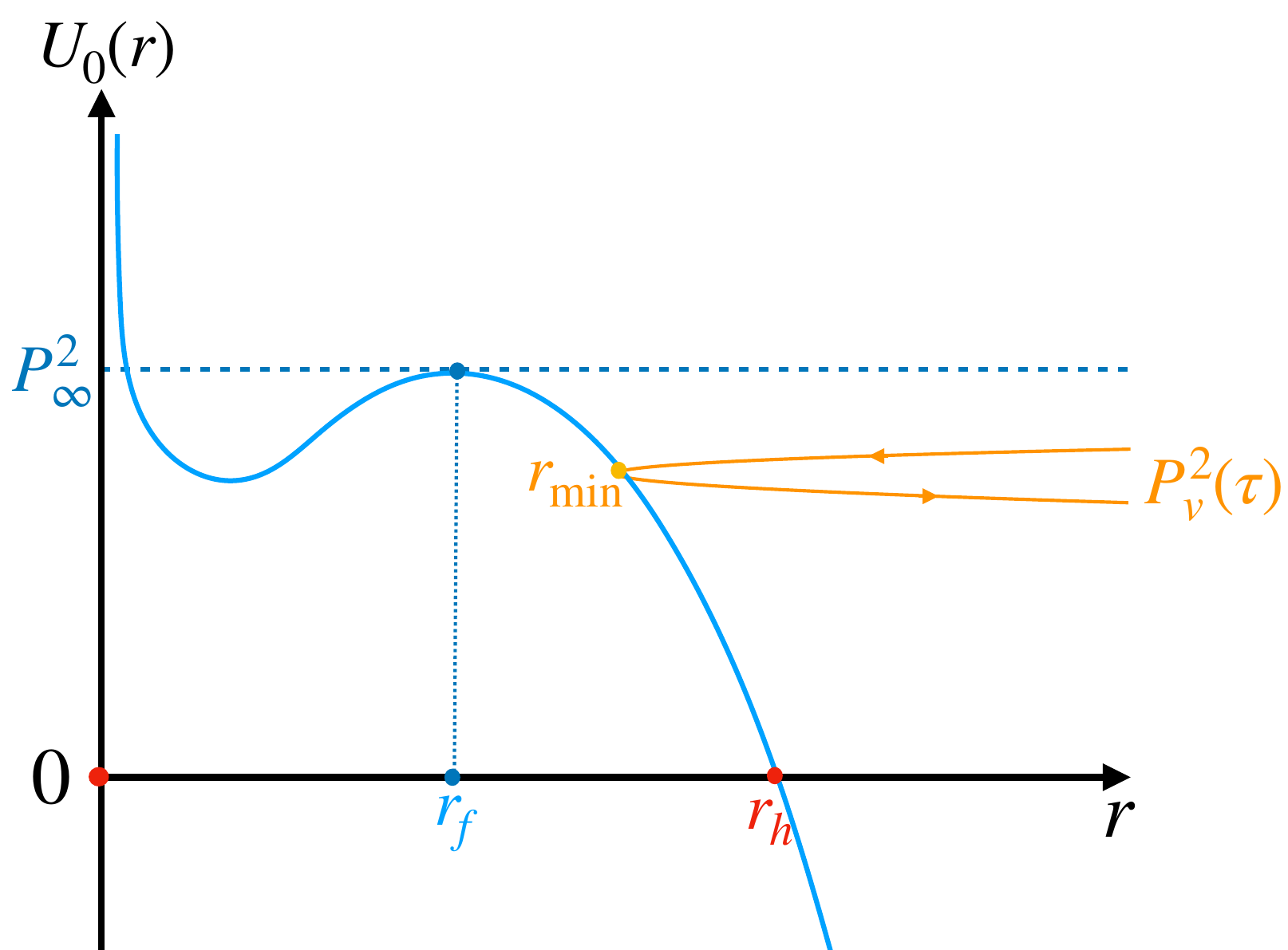}
	\includegraphics[width=3in]{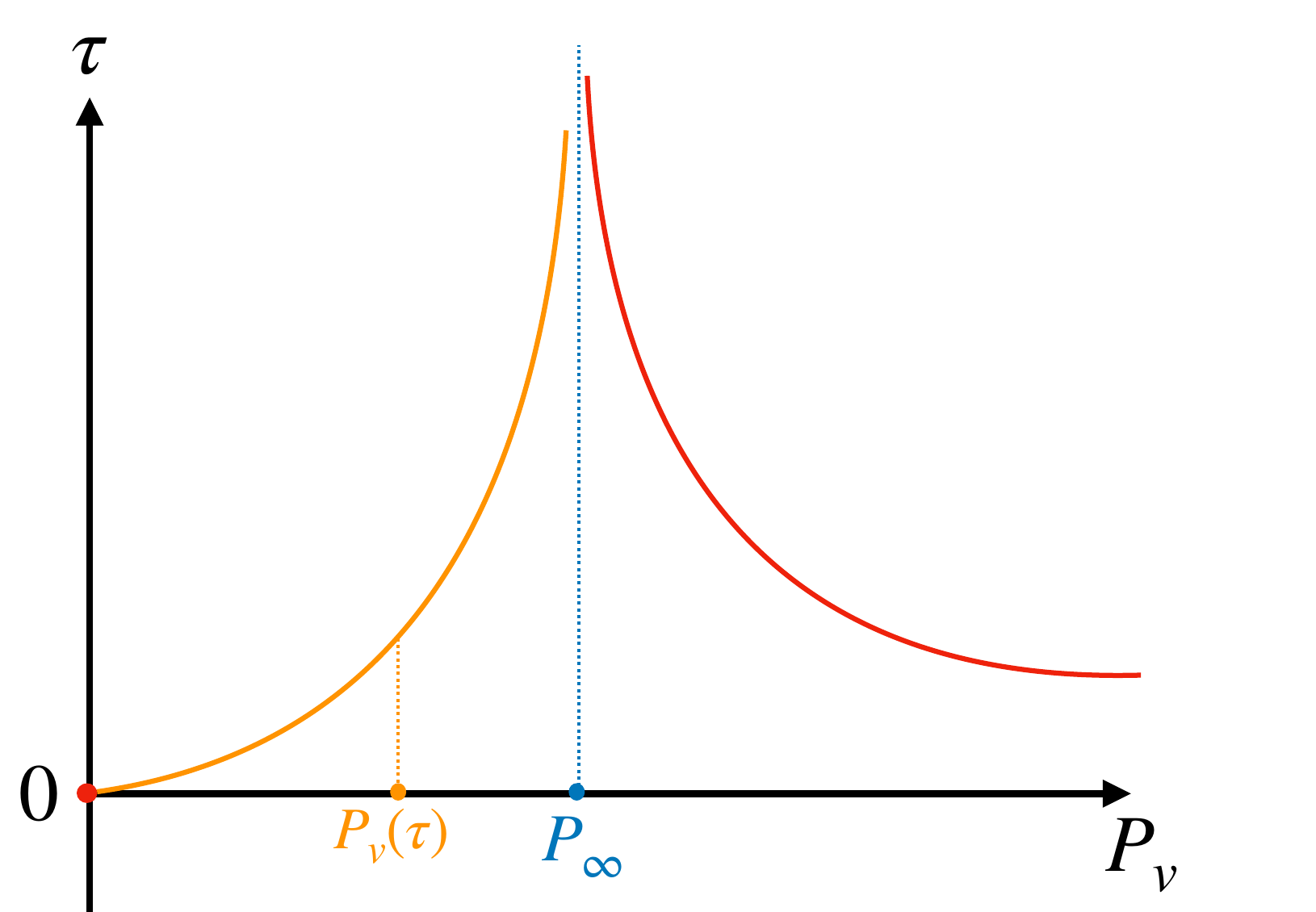}
	\caption{Left: A characteristic potential with a local maximum at $r=r_f$ and $U_0(r_f)= P^2_{\infty}$. Right: The relation between the conserved momentum $P_v$ and boundary time $\tau$.}\label{fig:Potential}
\end{figure}

The left panel in figure ~\ref{fig:Potential} illustrates a typical potential. Due to the factor $f(r)$ in eq.~\reef{potent}, the effective potential vanishes at the horizon $r=r_h$. We are interested in trajectories that begin and end at the asymptotic boundaries, \ie at $r\to\infty$, as illustrated in figure \ref{fig:TFDbalckhole}. For a given momentum $P_v$, the trajectory reverses direction at some finite radius, where the particle bounces off by the effective potential. That is, the extremal surface reaches the minimal radius $\rmin$, which is determined by $U(\rmin)= P_v^2$. More importantly, one can show that the time derivative of the observable \reef{eq:defineV} with respect to the boundary time $\tau$ is given by
\begin{equation}\label{eq:dVdt01}
	\frac{d \mathcal{C}_{\rm gen}}{d\tau} =\frac{\Omega_{k, d-1} L^{d-2}}{\GN}\,P_v(\tau)\,,
\end{equation}
because  the time evolution of the codimension-one surface is always extremal with respect to the functional $\mL_{\rm{gen}}$.

From eq.~\reef{eq:dVdt01}, it is straightforward to show that the linear growth at late times is due to the fact that the conserved momentum $P_v(\tau)$ at $\tau \to \infty $ approaches a fixed constant. However, note that the above expressions assume the existence of the extremal surface in the late-time limit $\tau \to \infty$. This fact is related to the condition that the effective potential $U_0(r)$ presents at least one local maximum inside the horizon. 
In order to see that, we can rewrite the relation between the boundary time and the conserved momentum as
\begin{equation}\label{eq:boundarytimetR02}
	\tau\equiv 2t_{\mt{R}}= -2\int^{\infty}_{r_{\rm{min}}} dr\, \frac{P_v}{f(r)\sqrt{P_v^2-{U}_0(r)}}  \,.
\end{equation}
Finding the extremal surface anchored on a specific boundary time slice $\tau$ thus corresponds to solving the Hamiltonian system \eqref{eq:Hamiltonian} with a given conserved momentum $P_v$ that is fixed by the boundary time. Now the integrand diverges at two points. The first is at the horizon where $f(r) \simeq f'(r_h)(r-r_h)$. However, we define the integral by the Cauchy principal value associated with this singularity, which is finite. The second divergence is at the turning point of the analogue particle where generically we have $U_0(r) \simeq P_v^2 + U'_0(r_{\rm{min}}) (r-r_{\rm{min}})$. This yields an integrable singularity and hence the boundary time remains finite. However, if the effective potential has a local maximum, we can tune $P_v\to P_\infty$ where $U'_0(r_{\rm{min}})\to 0$, \ie at the critical momentum, $U_0(r) \simeq P_\infty^2 + \frac12\, U''_0(r_f) (r-r_f)^2$ where $r_f$ is the critical value of $\rmin$. With this tuning,
the singularity at $\rmin$ in eq.~\reef{eq:boundarytimetR02} is no longer integrable and the boundary time diverges, as shown in the right panel of figure \ref{fig:Potential}. In other words, the existence of the extremal surface at late times is related to the local maximum of the effective potential. We can specify the local maximum at $r=r_f < r_h$ by 
\begin{equation}
	U_0(r_f)= P^2_{\infty}\,, \quad U_0'(r_f)=0\,,\quad  U_0''(r_f)\le 0 \,.
	\label{critical8}
\end{equation}
Finally, we can conclude that there is an infinite class of observables $\mC_{\rm{gen}}$ which exhibits linear growth at late times, \ie
\begin{equation} \label{laterate}
	\lim_{\tau \to \infty}  \(    \frac{d	\mC_{\rm gen}}{d\tau} \)  = \frac{\Omega_{k, d-1} L^{d-2}}{\GN} P_{\infty} \,.
\end{equation}

Above, we only considered the simplest approach where the same functional $F$ \eqref{eq:Ccodimensionone} is used to determine the extremal surface and to evaluate the observable. As already noted above, more generally, we can also consider gravitational observables of the form
\begin{equation}\label{eq:obsdef}
	O_{F_1,\Sigma_{F_2}}(\Scft) =\frac{1}{\GN L} \int_{\substack{\Sigma_{F_2}  }}\!\!\!\!\! d^d\sigma \,\sqrt{h} \,F_1(g_{
		\mu\nu}; X^{\mu}) \,,
\end{equation}
where the extremal surface $\Sigma_{F_2}$ is derived with respect to the scalar functional $F_2$ while the observable is evaluated with, \ie the integrand on the extremal surface, is given by an independent scalar function $F_1$. We can apply the same method introduced before to solve the extremal surfaces associated with the function $F_2$, \ie solving the classical equation of motion with a potential $U_2(r)$. However, the time derivative of $O_{F_1,\Sigma_{F_2}}$ would not be simply given by the conserved momentum $P_v$ that is defined by $ P_v^2=U_2(\rmin)$. Instead, we must reconsider the derivation of eq.~\reef{eq:dVdt01}.  In the present case, it is a more complicated integral term along the extremal surface: 
\begin{equation}\label{eq:dOdtau}
	\begin{split}
		\frac{dO_{F_1,\Sigma_{F_2}}}{d\tau} &= \frac{\Omega_{k, d-1} L^{d-2}}{\GN}\(  \sqrt{ \bar{U}_1}+
		 P_v \frac{d P_v}{d \tau} \int^{\infty}_{r_{\rm{min}}} dr\, \frac{ \sqrt{ U_1(r)U _2(r)} -\sqrt{\frac{\bar{U}_1}{\bar{U}_2}}U_2(r) }{f(r) \(P_v^2-U_2(r)\)^{3/2}}  \)\,,
	\end{split}
\end{equation}
where $U_1(r)$ is the effective potential that would be derived from $F_1$, and $\bar{U}_i = U_i \( \rmin \)$.  The non-vanishing bulk integral term reflects the fact that the surface $\Sigma_{F_2}$ is not extremal with respect to the integral $F_1$ when $F_1 \ne F_2$.  However, It has been demonstrated in \cite{Belin:2021bga} that the bulk integral terms in \eqref{eq:dOdtau} are suppressed at the late times due to the exponential decay of $d P_v /d \tau$. Consequently, we can conclude that  
the growth rate of $O_{F_1,\Sigma_{F_2}}$ at late times is dominated by the leading constant, \viz 
\begin{equation}
		\lim_{\tau \to \infty} \frac{dO_{F_1,\Sigma_{F_2}}}{d\tau}  = \frac{\Omega_{k, d-1} L^{d-2}}{\GN} \sqrt{U_1(r_f)}\,,\\
\end{equation}
where we note again that the final slice located at $r=r_f < r_h$ is determined by the scalar functional $F_2$ via $U_2(r_f) = P_\infty^2$ and $U_2'(r_f)=0$.

\subsection{Codimension-Zero Observables}\label{sec:generalization}

The analysis for codimension-zero observables defined in eq.~\eqref{eq:CgenZero} is similar to that above. The codimension-zero subregion is defined by two extremal surfaces with respect to the corresponding functionals. The key point in finding the extremal subregion is that one can rewrite the gravitational observables in terms of two boundary terms evaluated on $\Sigma_\pm$. As a result, the extremization for the extremal subregion $\mathcal{M}_{G,F_{\pm}}$ is equivalent to independently finding the two extremal hypersurfaces $\Sigma_{\pm}$. 

Taking the AdS black hole background \eqref{eq:AdSBH} as an explicit example, the codimension-zero functional $\mC_{\rm{gen}} $ defined in eq.~\eqref{eq:CgenZero} becomes
\begin{equation}\label{eq:defineV0}
	\mC_{\rm{gen}}(\tau)= \frac{\Omega_{k, d-1} L^{d-2}}{\GN}\sum_{\veps=+,-}\int_{\Sigma_\veps} d\sigma\, \[ \(\frac{r_\veps}{L}\)^{d-1}\sqrt{-f(r_\veps){\dot v_\veps}^2+2\dot v_\veps\,\dot r_\veps}\ a_\veps(r_\veps) - \veps \dot{v}_\veps\,b(r_\veps)\] \,,
\end{equation}
where a new function $b(r)$ arises in the two boundary integrals by integrating the bulk term by parts, \ie  $\sqrt{g}  \, G(g_{\mu\nu}) = G(r)  \( \frac{r}{L} \)^{d-1} \equiv L\,\frac{\partial b(r)}{\partial r} $.  Similar to the extremization problem described previously, we can identify two independent Lagrangians, \ie
\begin{equation}\label{eq:Lagragianpm}
	\mathcal{L}_\pm \equiv \(\frac{r}{L}\)^{d-1}\sqrt{-f(r){\dot v}^2+2\dot v\,\dot r}\ a_\pm(r) \mp \dot{v}\,b(r)
\end{equation}
for the two hypersurfaces $\Sigma_\pm$.  We will not reproduce the  detailed analysis here, but rather refer the interested reading to \cite{Belin:2022xmt}. The crucial point is that the profiles of $\Sigma_\pm$ are determined by two classical mechanics problems: 
\begin{equation}
0=	\dot{r}^2 + \mathcal{U}_\pm(P^\pm_v, r) \equiv  \dot{r}^2 + U_0(r)-(P^\pm_v\pm b(r))^2\,,
\end{equation}
where $U_0$ is defined as in eq.~\eqref{potent}, and the conserved momenta $P_v^\pm$ are given by
\begin{equation}\label{eq:moment}
	P^\pm_v=\frac{\partial {\cal L}_\pm}{\partial \dot{v}}=\dot r -f(r)\,\dot v \mp b(r)\,.
\end{equation}
The two conserved momenta also determine the growth rate of the codimension-zero extremal functional \eqref{eq:defineV0} as 
\begin{equation}\label{eq:dVdt}
		\frac{d}{d\tau}\,\mC_{\rm{gen}}(\tau)=  \frac{\Omega_{k, d-1} L^{d-2}}{\GN} \(  P_v^+(\tau)+P_v^-(\tau) \)\,.
\end{equation}
The linear growth at late times is also realized when the effective potential contains a local maximum inside the horizon, \ie 
\begin{equation}\label{critic}
	\mathcal{U}_\pm(P^\pm_{\infty}, r_{f,\pm}) = 0\,,\quad  \partial_r\,\mathcal{U}_\pm(P^\pm_{\infty}, r_{f,\pm}) = 0 \,, \quad  \partial^2_r\,\mathcal{U}_\pm(P^\pm_{\infty}, r_{f,\pm}) \le  0   \,. 
\end{equation}
The latter yields the extremal surfaces anchored to the boundaries at infinite time, and the corresponding $P_v^\pm (\tau)$ approach constants $P_{\infty}^\pm$ at late times.

Finally let us reiterate that the general complexity=anything proposal  \cite{Belin:2022xmt} involves  two pairs of bulk and boundary functionals, \ie the observable is evaluated with $(G_1, F_{1,\pm})$ and the extremal region is determined with $(G_2, F_{2,\pm})$. The generalized observables can be written as
\begin{equation}\label{eq:O1}
	\begin{split}
		&O\[G_1,F_{1,\pm},\mathcal{M}_{G_2,F_{2,\pm}}\] (\Scft)=\frac{1}{\GN L }\int_{ \Sigma_+[G_2,F_{2,+}]}\!\!\!\!\!\!\!d^d\sigma \,\sqrt{h} \,F_{1,+}(g_{\mu\nu}; X^{\mu}_+) \\
		&\qquad
		+\frac{1}{\GN L }\int_{   \Sigma_-[G_2,F_{2,-}]}\!\!\!\!\!\!\!d^d\sigma \,\sqrt{h} \,F_{1,-}(g_{\mu\nu}; X^{\mu}_-) +\frac{1}{G_N L^2 } \int_{\mathcal{M}_{G_2,F_{2,\pm}}}\!\!\!\!\!\!d^{d+1}x \,\sqrt{g} \ G_1(g_{
			\mu\nu})\,. 
	\end{split}
\end{equation}
Similar to the codimension-one case, it can be shown that these observables still yield linear late-time growth \cite{Belin:2022xmt}:
\begin{equation}\label{eq:dVdtlimit02}
	\lim_{\tau \to \infty}\frac{d}{d\tau}\, \Big(O\!\[G_1,F_{1,\pm}, \mathcal{M}_{G_2,F_{2,\pm}}\] \Big) = \frac{\Omega_{k, d-1} L^{d-2}}{\GN} \(  P_{\infty}^+(F_{1,+}, G_1) +P_{\infty}^-(F_{1,-}, G_1) \)\,.
\end{equation}
This expression involves two `fake' momenta
\begin{equation}
	P^\pm_{\infty}(F_{1,\pm}, G_1) \equiv  \sqrt{ U_1(r_{f,\pm})} \mp b_1(r_{f,\pm}) \,.
\end{equation}
with the final slices at $r=r_{f,\pm}$ are determined by the effective potentials constructed from $(G_2, F_{2,\pm})$.

%%%%%%%%%%%%%%%%%%%%%%%%%%%%%%%%%%%%%%
\section{Complexity = Anything Revisited}\label{sec:end}
%%%%%%%%%%%%%%%%%%%%%%%%%%%%%%%%%%%%%%

As reviewed in the previous section, both the codimension-one observables in eqs.~\eqref{eq:Ccodimensionone} and \eqref{eq:obsdef} and the codimension-zero observables in eqs.~\eqref{eq:CgenZero} and \eqref{eq:O1} exhibit linear growth at late times. This behaviour is related to the existence of a local maximum in the corresponding effective potential, which defines a final constant-radius slice at $r=r_f$. Intuitively, the late-time linear growth arises from the corresponding extremal surfaces expanding out along this final slice. While these new gravitational observables offer fresh insight into the geometry of the black hole interior, they are typically only probing a portion of the interior geometry since the final slice at $r=r_f$ also acts as a barrier preventing the extremal surfaces from reaching the singularity. 

In order to probe the geometry of the singularity, one can push the final slice closer to the singularity by tuning the various couplings that appear in gravitational observables. We discuss this approach with a particular example in the next section -- see also section \ref{sec:disc}. Instead, here, we turn to a puzzle which first appeared in \cite{Belin:2021bga}. It was found that a particular codimension-one observable with a $C^2$ term, \ie the square of Weyl tensor, only yields extremal surfaces at late times for a limited range of the corresponding coupling. That is, the desired local maximum in the effective potential disappears beyond a limited range of the coupling. In this context, no choice of the coupling pushes $r_f$ close to the singularity. 

However, with a more careful examination, we will show that the resolution of this puzzle is that beyond the `allowed' range of the coupling, the surfaces yielding the maximal value of the observable are pushed to the edge of the allowed phase space. Hence these `maximal' surfaces are no longer locally extremal (everywhere). That is, they are not found by solving the equations derived from extremizing the observable, as described in section \reef{recap}. Further, in certain instances, the maximal surfaces hug the black hole singularity.

%%%%%%%%%%%%%%%%%%%%%%%%%%%%%%%%%%%%%%%%%%
%%%%%%%%%%%%%%%%%%%%%%%%%%%%%%%%%%%%%%%%%%
\begin{figure}[ht!]
	\centering
	\includegraphics[width=2.8in]{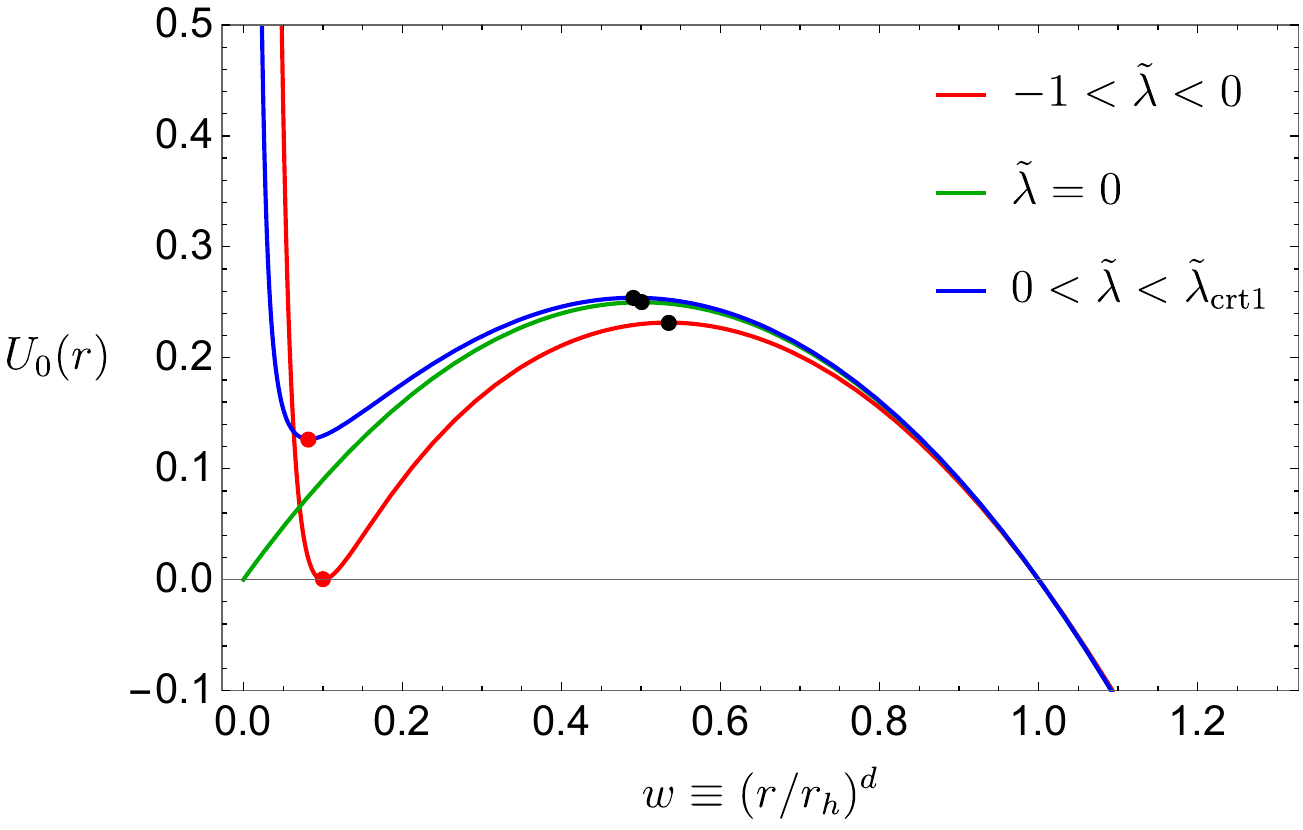}\qquad 
	\includegraphics[width=2.8in]{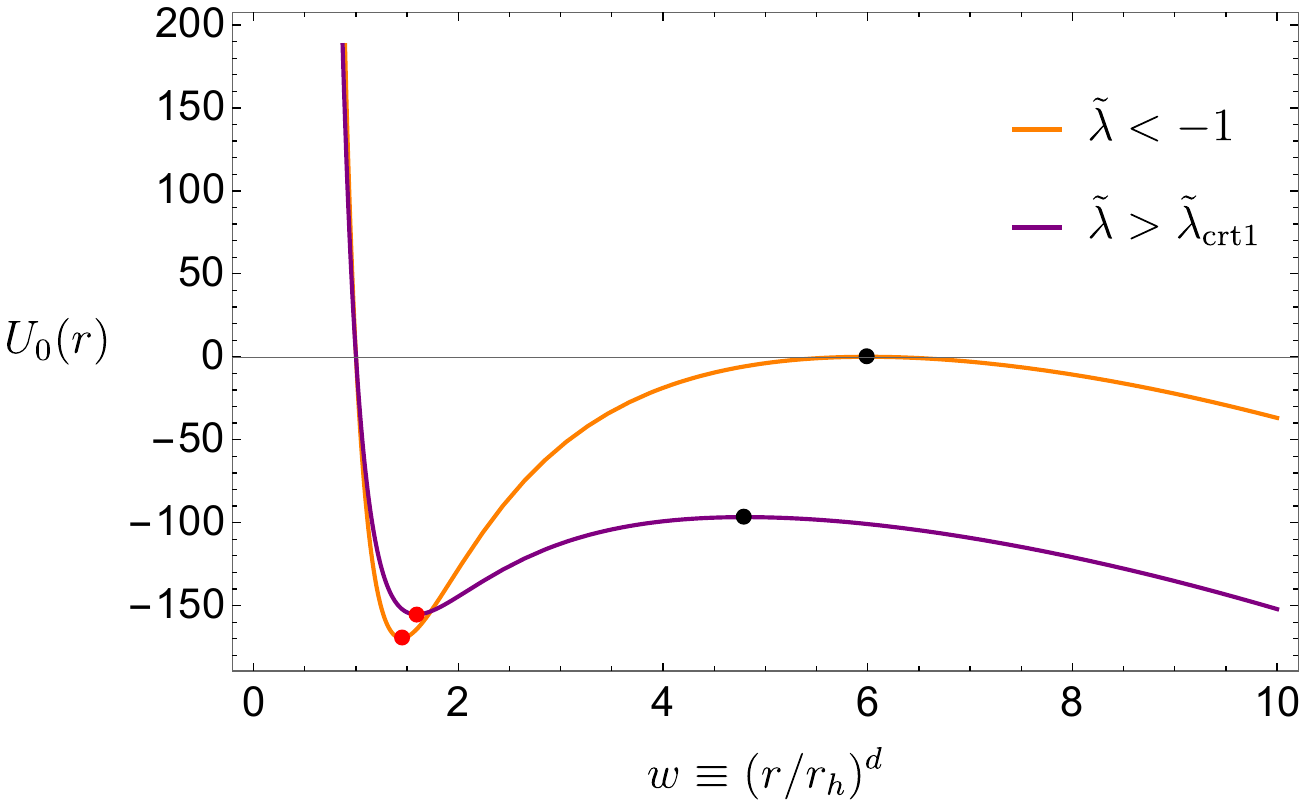}
	\caption{Left: The effective potentials $U_0$ with $-1<\tilde{\lambda} < \tilde{\lambda}_{\rm{crt}1}$. Right: The effective potentials are shown for $\tilde{\lambda} <-1$ or $\tilde{\lambda} > \tilde{\lambda}_{\rm{crt}1}$. The black and red dots represent the local maximum and local minimum of the effective potential, respectively. The late-time linear growth of $\mC_{\rm gen}$ is associated with the existence of a (positive) local maximum located inside the black hole horizon (\ie taking $-1<\tilde{\lambda} < \tilde{\lambda}_{\rm{crt}1}$). }
	\label{fig:potentialC2}
\end{figure}
%%%%%%%%%%%%%%%%%%%%%
%%%%%%%%%%%%%%%%%%%%%

Let us begin then by considering an explicit example of the codimension-one observables \reef{eq:Ccodimensionone}, \ie 
\begin{equation}\label{eq:generalziedC2}
	\mathcal{C}_{\rm gen}= \frac{1}{\GN L} \int d^d\sigma \,\sqrt{h_{ij}}\left(1 + \lambda \,L^4\,C^2 \right)\,, 
	%\quad \quad F_1=1 + \lambda \,L^4\,C^2 \,,
\end{equation}
where the second term is proportional to the square of the Weyl tensor, $C^2= C_{\mu\nu\rho\sigma}\,C^{\mu\nu\rho\sigma}$. The strength of this higher curvature term is controlled by the dimensionless coupling $\lambda$. This explicit example was carefully examined in appendix A of both \cite{Belin:2021bga,Belin:2022xmt}. See also \cite{Omidi:2022whq,Wang:2023eep} for the studies of this functional \eqref{eq:generalziedC2} in other spacetime backgrounds. We can proceed to evaluate the profile of the extremal surfaces as described above in section \ref{codone}.
In particular, the radial profile is determined by the classical mechanics system described by eq.~\reef{eq:Hamiltonian} with the effective potential defined in eq.~\reef{potent}. For simplicity, let us consider the planar vacuum black holes for which the blackening factor $f(r)$ is given by eq.~\reef{blacken} with $k=0$. Then the factor $a(r)$ associated with the Weyl-squared observable above becomes
\begin{equation}\label{eq:CCar}
	a(r) =1 + \tilde{\lambda}\,   \frac{L^4 \omega^{2(d-2)}}{r^{2d}} =1 + \tilde{\lambda}\, \(  \frac{r_h}{r} \)^{2d}\,, 
\end{equation}
with $\tilde{\lambda} =d(d-1)^2(d-2)\,\lambda$. The corresponding effective potential \reef{potent} is then conveniently written as
\begin{equation}\label{eq:ham3}
		U_0(r)=\(\frac{r_h}{L}\)^{2d}\left(w-w^2\right)\left(
		1 + \frac{\tilde{\lambda}}{w^2}\right)^2 \,,
\end{equation}
by using the dimensionless radial coordinate $w= (r/r_h)^d$. In figure \ref{fig:potentialC2}, we show some characteristic plots for the effective potential $U_0(r)$ with various $\tilde{\lambda}$.  We remark that the effective potential or the factor $a(r)$ are always divergent at $r=0$ for any nonzero value of $\tlam$ because $C^2$ diverges at the black hole singularity. 

Following the discussion in section \ref{recap}, the late-time growth is determined by the critical points in the potential $U_0$. In particular, we are looking for positive maxima within the black hole horizon, as shown in eq.~\reef{critical8}. Examining the potential in eq.~\reef{eq:ham3}, there can be a single positive maximum $w_f= (r_f/r_h)^d$, which occurs behind the horizon, \ie $0<w_f<1$. However, as explained in \cite{Belin:2021bga}, this maximum only occurs when the $C^2$ coupling satisfies\footnote{The effective potential will also have a local maximum for $\tilde{\lambda}<-1$ and $\tilde{\lambda} > \tilde{\lambda}_{\mathrm{crt}2}= \frac{1}{8} (47+13\sqrt{13})$. However, the $U_0$ is zero at the maximum in the first range and negative, in the second. Further in both cases, the maxima occur outside of the horizon.}
\begin{equation}\label{range1}
	-1<\tilde{\lambda} < \tilde{\lambda}_{\mathrm{crt}1}\equiv \frac{1}{8} (47-13\sqrt{13})\,.
\end{equation}

\begin{figure}[t]
	\centering
	\includegraphics[height=1.6in]{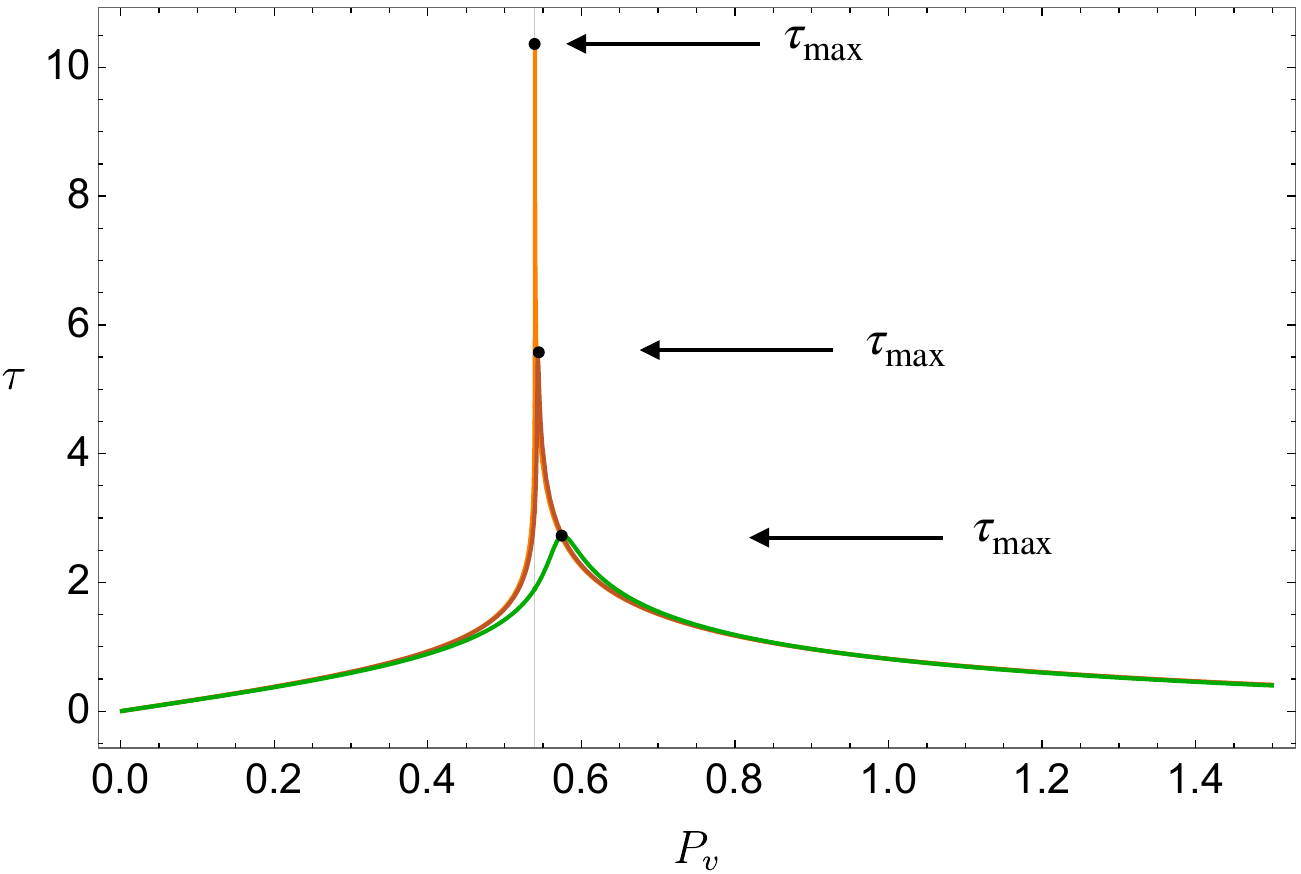}
	\hspace{.2in}
	\includegraphics[height=1.6in]{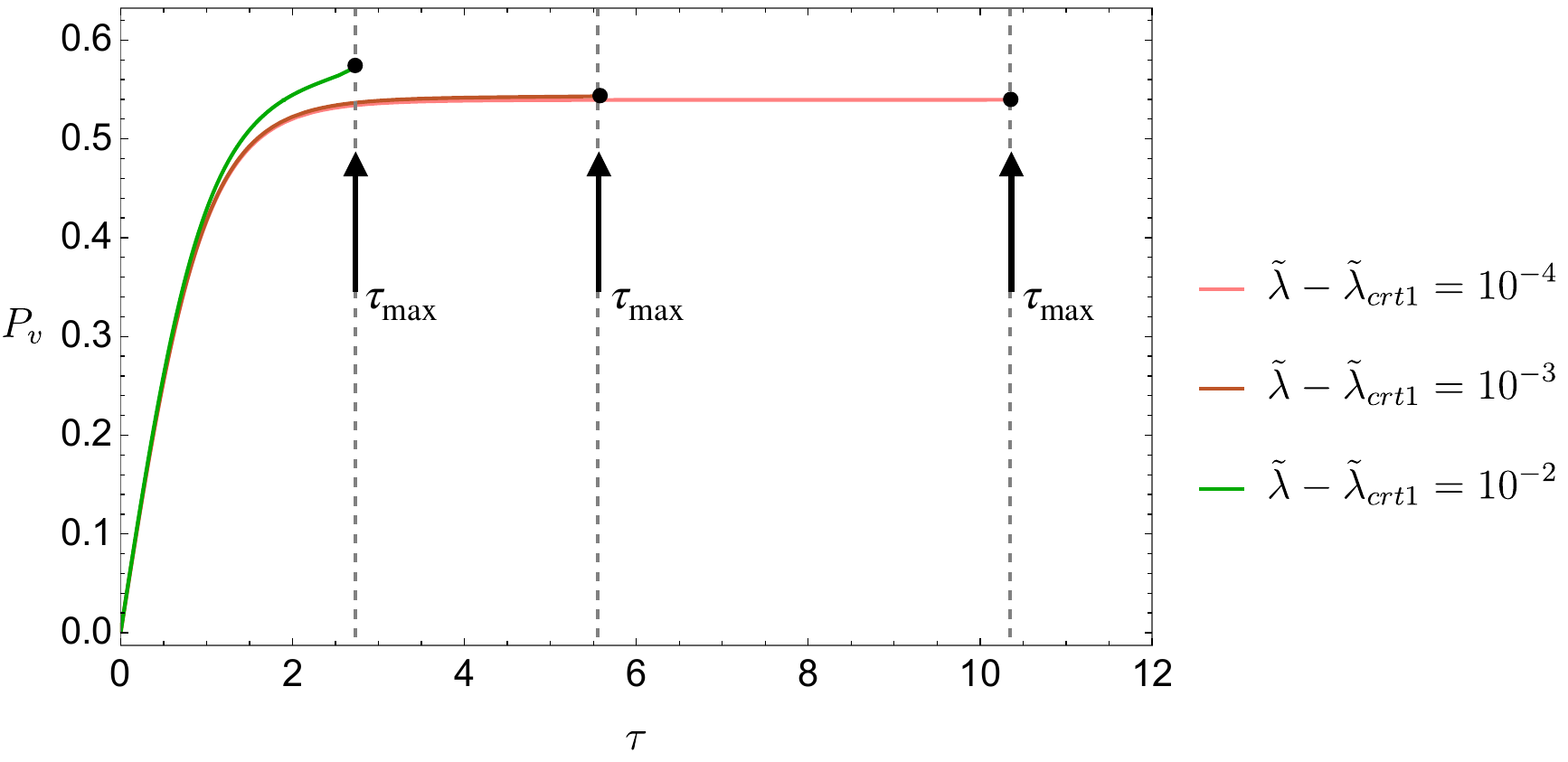}
	\caption{Left: The boundary time $\tau$ as a function of the conserved momentum $P_v$, when the coupling $\tilde{\lambda}$ (slightly) exceeds the critical value $\tilde{\lambda}_{\mathrm{crt}1}$. Note that the pole corresponding to $\tau\to\infty$ in the right panel of figure \ref{fig:Potential} is replaced by a finite maximum $\tau_{\rm max}$. Right: The corresponding growth rate of $\mathcal{C}_{\rm{gen}}$ as a function of the boundary time $\tau$. Recall $P_v\propto d \mathcal{C}_{\rm gen}/d\tau$ from eq.~\reef{eq:dVdt01}.}\label{fig:Pvtau}
\end{figure}  

Now one may ask what happens to the time evolution of the observable \reef{eq:generalziedC2} when the coupling lies outside of the range given above. In particular, in the left panel of figure \ref{fig:Pvtau}, we consider the plot of the boundary time $\tau$ as a function of the conserved momentum $P_v$ for $\tilde{\lambda}\gtrsim\tilde{\lambda}_{\mathrm{crt}1}$. For couplings in the allowed range \reef{range1}, the corresponding plot is shown in the right panel of figure \ref{fig:Potential} and recall that there is a pole at $P_v=P_\infty$ corresponding to $\tau\to\infty$. Instead in figure \ref{fig:Pvtau}, this pole is replaced by a finite peak and so the boundary time seems to reach a maximum $\tau_{\rm max}$. 
Further we can tune $\tilde{\lambda}-\tilde{\lambda}_{\mathrm{crt}1}\ll 1$ to make $\tau_{\rm max}$ arbitrarily large. 
Plotting the same curve (or rather the portion up to $\tau_{\rm max}$) but with $P_v$ as a function of $\tau$, as shown in the right panel of figure \ref{fig:Pvtau}, we gain some insight into the time evolution of our observable since $d\mathcal{C}_{\rm{gen}}/d\tau$ is proportional to $P_v$ -- see eq.~\reef{eq:dVdt01}. However,  considering the case $\tilde{\lambda}-\tilde{\lambda}_{\mathrm{crt}1}=10^{-4}$, we see that $\mathcal{C}_{\mt{gen}}$ begins to grow with the growth rate quickly becoming constant. That is, as in the allowed range, we rapidly reach a phase of linear growth with $\tau$, however, this phase extends for a finite period ending at $\tau=\tau_{\rm max}$. After that time, the saddle point (\ie the extremal surface) no longer exists and we do not have a value for the observable or the growth rate. We also see that for larger values of $\tilde{\lambda}-\tilde{\lambda}_{\mathrm{crt}1}$, $\tau_{\rm max}$ quickly decreases and the phase of linear growth disappears.

\begin{figure}[t]
	\centering
   \includegraphics[width=2.5in]{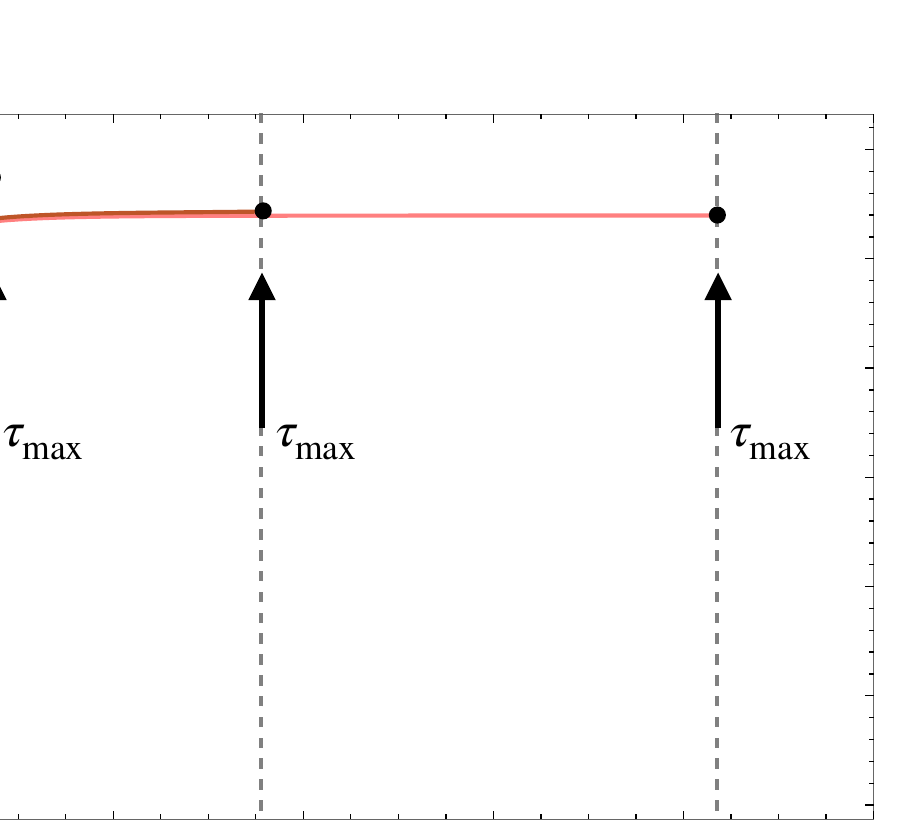}
	\includegraphics[width=3.2in]{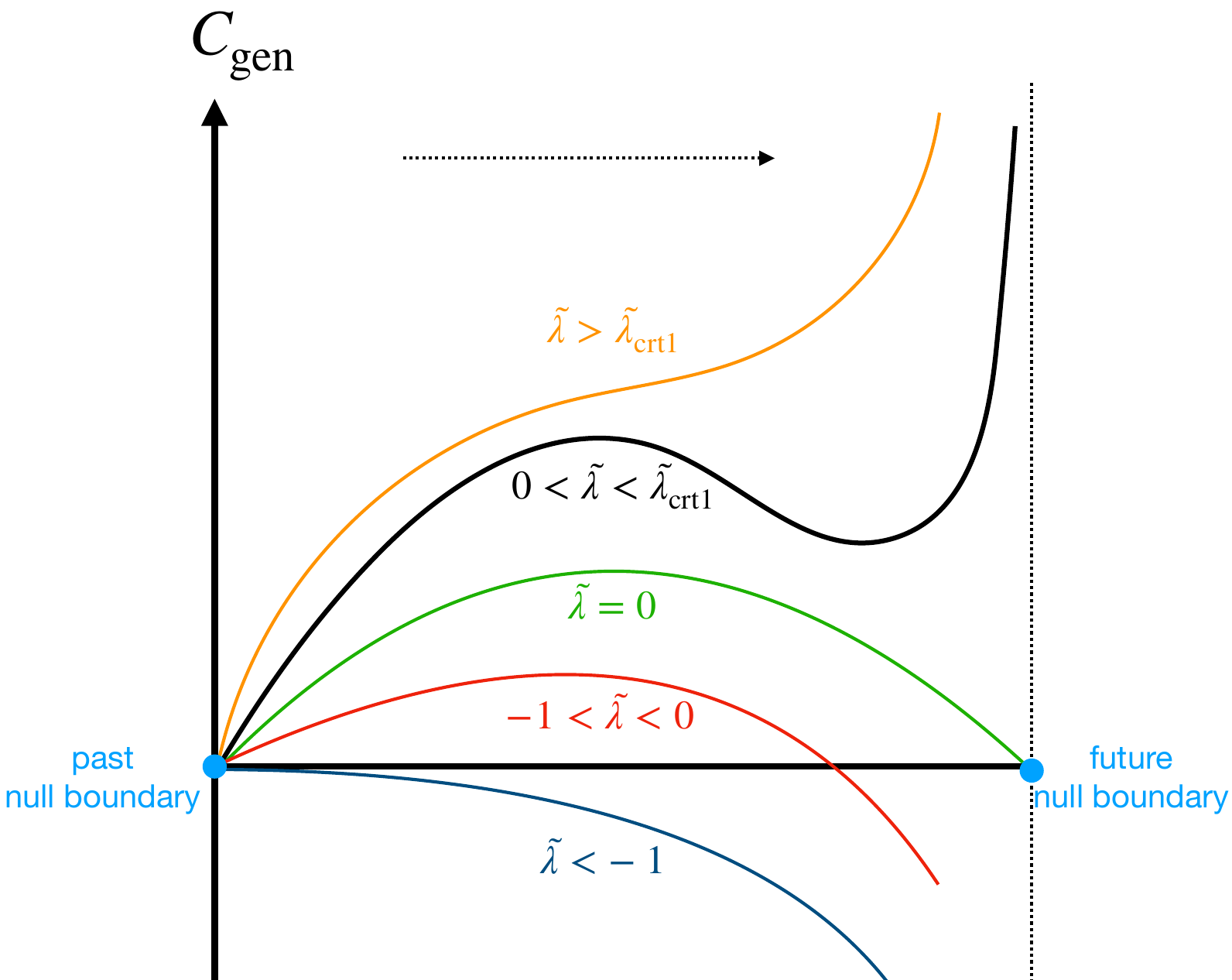}
	\caption{Left: All possible spacelike hypersurfaces connecting the fixed time slices on the left and right asymptotic boundaries fill the WDW patch, as indicated by the blue-shaded region in the Penrose diagram. We consider a one-parameter family of smooth candidate surfaces that sweep across the full WDW patch. Right: The value of ${\mathcal C}_\mt{gen}$ evaluated on the candidate surfaces for various regimes of the coupling $\tlam$. Moving from the left to the right on the horizontal axis corresponds to gradually sweeping from the past null boundary to the future boundary of the WDW patch, which includes the spacelike singularity. }\label{fig:tall1}
\end{figure}

This result is somewhat disconcerting and so we examine the extremal surfaces from a fresh perspective with figure \ref{fig:tall1}.   In principle, we can evaluate the generalized complexity \reef{eq:generalziedC2} for any spacelike surface connected to the boundary time slice, which we choose with large $\tau$. As shown in the left panel, these surfaces will always lie within the corresponding WDW patch. Of course, the full family of these surfaces is infinite-dimensional, however, to sketch the characteristic behaviour we consider a one-parameter family of smooth candidate surfaces that sweep across the full WDW patch, as illustrated in the figure. To be concrete, we could let these be constant mean curvature surfaces, \ie surfaces on which $K=$constant, as appear in section \ref{sec:CMC}. Now irrespective of the value of the coupling, as the surfaces approach the past boundary of the WDW patch, $\mathcal{C}_{\rm{gen}}\to0$ because the boundary surfaces are null while $C^2$ remains finite there. In contrast, approaching the future boundary yields $\mathcal{C}_{\rm{gen}}\to\pm \infty$ for positive and negative $\tilde\lambda$, respectively, because $\sqrt{h}\,C^2\propto 1/r^{2d-1}$ on constant $r$ surfaces near the singularity for the planar vacuum black holes.\footnote{See comments on UV divergences below.}

The right panel in figure \ref{fig:tall1} illustrates the expected behaviour of $\mathcal{C}_{\rm{gen}}$ between these two limits for four different choices of the coupling $\tilde\lambda$: {\it (i)} For positive $\tilde\lambda$ with $\tilde\lambda<\tilde{\lambda}_{\mathrm{crt}1}$, $\mathcal{C}_{\rm{gen}}$ rises from zero on the past boundary\footnote{Throughout this discussion, we are assuming that the boundary time $\tau$ is sufficiently large that the past boundary of the WDW patch does not touch the past singularity in the white hole region of the maximally extended Penrose diagram, as shown in the left panel of figure \ref{fig:tall1}.}  and reaches a local maximum when the candidate surface approximates the extremal surface. Next, $\mathcal{C}_{\rm{gen}}$ decreases to a local minimum when the minimum radius of the candidate surface falls in the vicinity of $w=w_{\rm min}$, the position of the local minimum in $U_0$ -- see figure \ref{fig:potentialC2}. Finally, $\mathcal{C}_{\rm{gen}}\to+ \infty$ as the surfaces approach the future boundary of the WDW patch. 
{\it (ii)} For positive $\tilde\lambda$ with $\tilde\lambda>\tilde{\lambda}_{\mathrm{crt}1}$ and $\tau>\tau_{\rm max}$, the critical points in effective potential $U_0$ have merged and become complex and so too, the critical points in the previous plot have disappeared. That is, $\mathcal{C}_{\rm{gen}}$ simply rises monotonically from zero on the past boundary of the WDW patch to $+\infty$ on the future boundary. {\it (iii)} For negative $\tilde\lambda$ with $\tilde\lambda>-1$, $\mathcal{C}_{\rm{gen}}$ rises from zero on the past boundary and reaches a local maximum, as in the first case. Next, $\mathcal{C}_{\rm{gen}}$ decreases with $\mathcal{C}_{\rm{gen}}\to- \infty$ as the surfaces approach the future boundary of the WDW patch. The curve may show some structure when the minimum radius of the candidate surface reaches near $w=w_{\rm min}$, but at best this would be a point of inflection. {\it (iv)} Finally, for negative $\tilde\lambda$ with $\tilde\lambda<-1$ and $\tau>\tau_{\rm max}$, the critical points in effective potential $U_0$ and the critical surface have disappeared. Hence $\mathcal{C}_{\rm{gen}}$ simply decreases monotonically from zero to $-\infty$ as the candidate surfaces sweep between the past and future boundaries.

For couplings outside of the allowed range \reef{range1} (\ie cases {\it (ii)} and {\it (iv)} above), we concluded that there are no (locally) extremal surfaces.  However, the discussion above argues that the surfaces which yield the maximal value of $\mathcal{C}_{\rm{gen}}$ are pushed to the boundary of the phase space of the allowed surfaces. In fact, the discussion for $\tlam<-1$ must be refined and we return to this question below in section \ref{sec:negative}. The correct result for $\tilde\lambda>\tilde{\lambda}_{\mathrm{crt}1}$  is that the `maximal' surface coincides with the future boundary of the WDW patch, \ie it consists of two null sheets extending from the boundary time slices $t_{\mt{R}}=t_{\mt{L}}=\tau/2$ to the future singularity and it hugs the singularity in between. In fact, the value of the observable diverges for these surfaces. We have some comments about regulating the calculation below in section \ref{regular1}.

However, this result raises several questions. For example, is the extremal surface found at early times (\ie $\tau<\tau_{\rm max}$) the correct saddle point? The above discussion reminds us that the WDW patch will always touch future singularity in the vacuum AdS black holes (for $\tau>0$), \eg see \cite{Carmi:2017jqz}. Hence the future boundary will always yield a (positive) divergent result for $\mathcal{C}_{\rm{gen}}$ and this would always be the maximal surface rather than the locally extremal surface. In fact, the same result applies to any positive coupling even with $0<\tilde\lambda<\tilde{\lambda}_{\mathrm{crt}1}$. Further, this would apply for any observable where $a(r\to0)\to+\infty$ irrespective of the structure of the effective potential \reef{potent}, \ie even if there are a number of extremal points between the singularity and the horizon. Of course, this means that as they stand such observables will not be very useful in diagnosing the interiors of vacuum AdS black holes. They may still yield finite results for other kinds of black holes, \eg carrying an electric charge. Further, one can `regulate' such observables to yield sensible finite results even in the vacuum case, as described in the next section.

\subsection{Regularization of $\mathcal{C}_{\mathrm{gen}}$} \label{regular1}

The preceding discussion is very heuristic. For example, evaluating the observable for all of the candidate surfaces yields UV divergences from the asymptotic regions, \eg see \cite{Carmi:2016wjl}. An implicit assumption then is that these UV divergences are identical for all of the candidate surfaces so that they can be ignored in comparing $\mathcal{C}_{\mathrm{gen}}$ for different surfaces. This would require some specific tuning of how the candidate surfaces approach the asymptotic boundaries. However, in this section, we demonstrate that we can use our standard analysis \cite{Belin:2021bga,Belin:2022xmt} to reach the same conclusions as above by introducing a regulated version of the observable \reef{eq:generalziedC2}.  For the regime $\tilde\lambda>0$, we proceed as 
follows:\footnote{We return to the case of $\tlam<-1$ in the next subsection.} Above we identified the source of the issue, \ie the maximal surface being pushed to the future boundary of the WDW patch, as $a(r\to0)\to+\infty$. The latter divergence can be ameliorated by adding a higher curvature term to the integrand of the observable with a small negative coefficient. For example, let us consider
\begin{equation}
	\mathcal{C}_{\mathrm{gen,reg}}=\frac{1}{\GN L} \int d^{d} \sigma \sqrt{h}\left( 1+\lambda\, L^{4} C^{2} - \lambda_4\, L^{8} C^{4}\right) \,,
	\label{observe4}
\end{equation}
with the Weyl square term  $C^2$ and a `subleading' term $C^4 \equiv \(C_{\mu\nu\alpha\beta} C^{\mu\nu\alpha\beta} \)^2$. Of course, we recover eq.~\reef{eq:generalziedC2} if we set $\lambda_4=0$. Here we consider $0<\lambda_4\ll1$ so that the new $C^4$ term has a minimal effect on $a(r)$ and the effective potential except where $r$ is very small. Then the new term will dominate so that $a(r\to0)\to-\infty$ for our regulated observable.

\begin{figure}[t]
	\centering
	\includegraphics[width=4in]{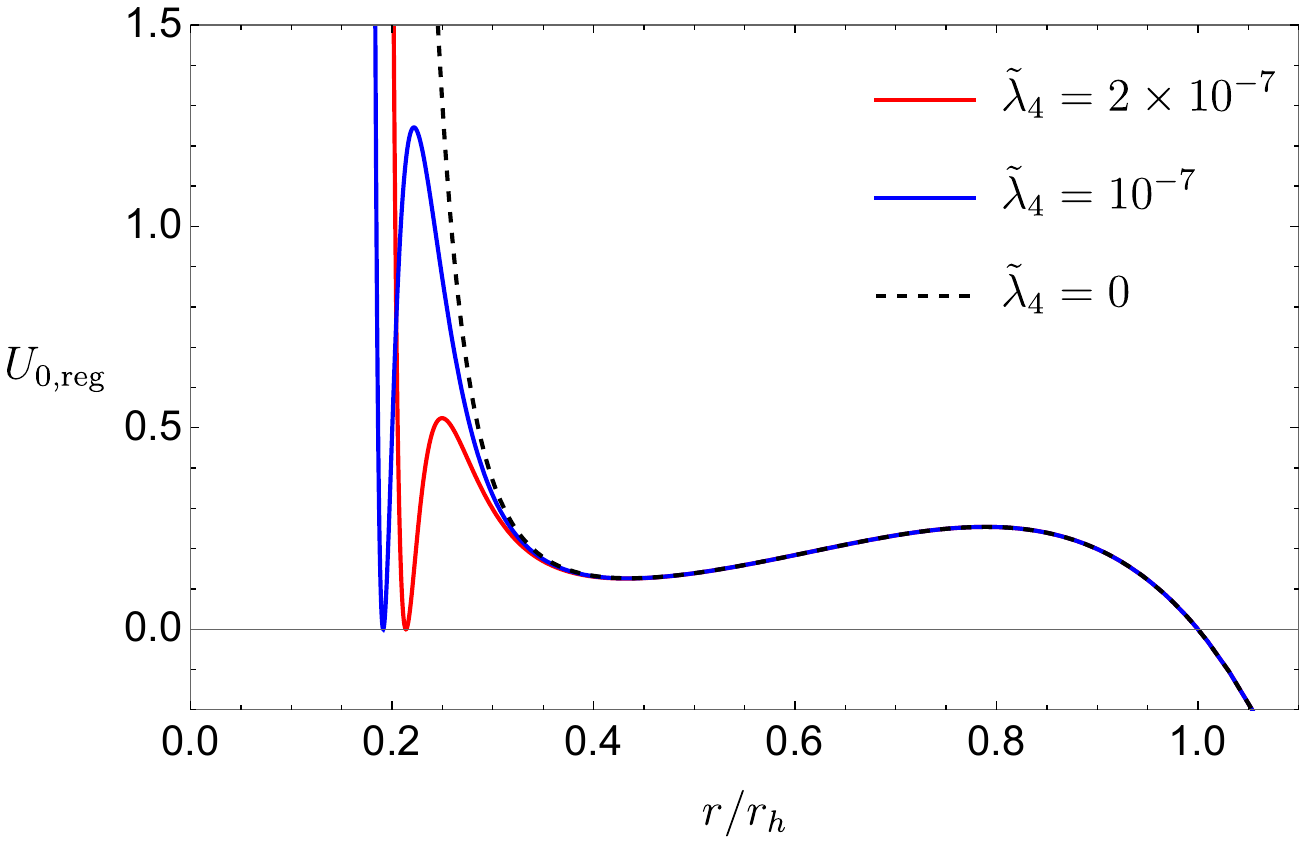}
	\caption{The effective potentials associated with regulated observable $\mathcal{C}_{\mathrm{gen,reg}}$ in eq.~\reef{observe4}. In each of these cases, the $C^2$ coupling is fixed to be $\tlam=1/500 < \tlam_{\rm crt1}$.   }\label{fig:C4potential}
\end{figure}

To be precise, the factor $a(r)$ associated with the observable in eq.~\reef{observe4} becomes
\begin{equation} \label{CCfour}
a(r)= 1 +\tilde{ \lambda}  \( \frac{r_h}{r} \)^{2d}- \tilde{\lambda}_4 \( \frac{r_h}{r} \)^{4d} \,,
\end{equation}
where $\tilde{\lambda}$ is defined below eq.~\reef{eq:CCar} and $\tilde{\lambda}_4 =d^2(d-1)^4(d-2)^2\,\lambda_4$. The final term only comes into play when the last two terms are comparable, \ie $r\lesssim (\tlam_4/\tlam)^{\frac1{2d}}\, r_h$.
Given eq.~\reef{CCfour}, the effective potential in eq.~\reef{eq:ham3} is replaced by
\begin{equation}\label{eq:ham4}
		U_{0,\mt{reg}}(r)=\(\frac{r_h}{L}\)^{2d}\left(w-w^2\right)\left(
		1 + \frac{\tilde{\lambda}}{w^2}- \frac{\tilde{\lambda}_4}{w^4}\right)^2 \,.
\end{equation}
where $w= (r/r_h)^d$, as before. In figure \ref{fig:C4potential}, we show some characteristic plots for the effective potential $U_0(r)$ with various values of $\tilde{\lambda}_4$. In the regime $0<\tlam_4\ll1$ (and $\tlam>0$), it is straightforward to show that this potential has a global maximum at 
\begin{equation}\label{peak}
	w_f \simeq \sqrt{\frac{7\tlam_4}{3\tlam}}\qquad \longrightarrow
	\quad r_f \simeq \left(\frac{7\tlam_4}{3\tlam}\right)^{\frac1{2d}}\,r_h\,.
\end{equation}

As explained in appendix A of \cite{Belin:2022xmt}, this global maximum controls the linear growth at late times. Combining eqs.~\reef{blacken}, \reef{mass}, \reef{critical8} and \reef{laterate}, we find the late-time growth rate is given by
\begin{equation}	\label{laterate2}
	\lim_{\tau \to \infty}  \(    \frac{d	\mC_{\rm gen}}{d\tau} \)  = \frac{64\pi}{7\,(d-1)}\,\left(\frac{3\tlam}{7\tlam_4}\right)^{\frac34}\,\tlam\,M\,.
\end{equation}
Hence we have that as $\tlam_4\to0$, $r_f\sim \tlam_4^{1/2d}\to 0$ and the time rate of change diverges as $\tlam_4^{-3/4}$. That is, in this limit where we recover the original observable \reef{eq:generalziedC2}, the extremal surface approaches the singularity and becomes the future boundary of the WDW patch when $\tlam_4=0$. Similarly, we see that the observable also diverges as expected in this limit. Hence we have recovered the same results for which we argued in a more qualitative way above. 

While we have examined a specific example above, this kind of regularization is quite general. That is, given an observable for which $a(r\to0)\to+\infty$, we can introduce an additional higher curvature term to the integrand of the observable with a small negative coefficient to ensure that $a_\mt{reg}(r\to0)\to-\infty$. This ensures that the effective potential has a global maximum very close to the singularity at $r=0$, which controls the late-time evolution of the regulated observable. Furthermore, let us note that when we keep the regulator coupling (\eg $\tlam_4$) small but finite, the regulated observable yields finite results and is useful in probing the spacetime geometry in the vicinity of the singularity. For example, the speed with which the late-time growth rate diverges as the regulator coupling approaches zero should characterize the curvature divergence at the singularity -- see further discussion in section \ref{sec:disc}. Hence eq.~\reef{observe4} provides an example of an observable where tuning the parameters pushes the final $r=r_f$ slice near the singularity. We pursue this idea further with a slightly different (and simpler) approach below in section \ref{sec:CMC}.

\subsection{Negative Coupling $\tlam<-1$} \label{sec:negative}
Finally, let us consider the regime of the dimensionless parameter where $\tilde{\lambda} < -1$. As we explicitly calculated before, the corresponding potential $U_0(r)$ does not present any local maximum inside the horizon. On the contrary, it is straightforward to see that the effective potential defined in eq.~\eqref{eq:ham3} instead has a local maximum at
\begin{equation}\label{eq:criticalr}
w= w_{\rm crt} \equiv  \( \frac{r_{\rm crt}}{r_h} \)^d = \sqrt{-\tl} >1\,.
	\end{equation}
where the potential vanishes and which is always outside the horizon when $\tilde{\lambda} < -1$.\footnote{For $-1<\tlam<0$, $r=r_\mt{crt}$ is a spacelike surface inside the horizon and inside the late time surface $r=r_f$. Hence it does not play a role in finding the extremal surface. } Of course, one can also find a local minimum between the horizon and the local maximum, as illustrated in figure \ref{fig:potentialC2}.

\begin{figure}[t!]
	\centering
	\includegraphics[width=3.5in]{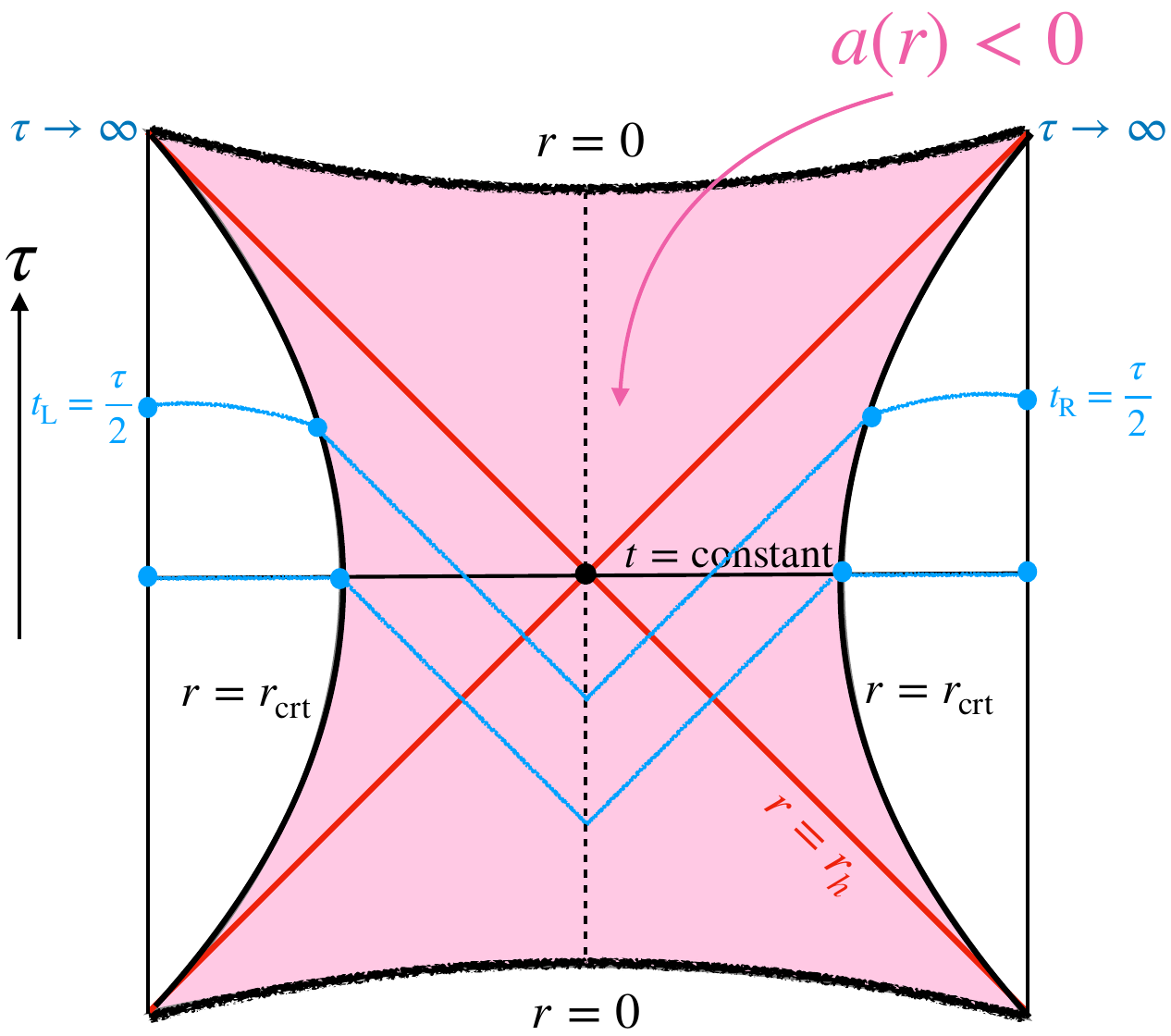}
	\caption{The pink region illustrates a part of the spacetime where the integrand of the observable in eq.~{eq:generlizedCV} is negative when $\tlam<-1$. That is, $a(r)<0$ inside the critical radius $r_{\rm crt} = (-\tl)^{1/2d}\, r_h$, which is defined in eq.~\eqref{eq:criticalr}. The surfaces that maximize the observable are shown in blue.}\label{fig:tworegions}
\end{figure}

It is obvious that our previous proof for the linear growth can not apply to the current case with $\tilde{\lambda} < -1$ due to the absence of a local maximum inside the horizon. This is a direct result of the absence of a smooth extremal surface connecting the left and right boundaries at late times.
Another noteworthy aspect of $\tilde{\lambda} < -1$ is that the volume measure, as represented by the integrand of $\mC_\mt{gen}$, becomes negative behind the critical radius since
\begin{equation}
	 a(r) < 0 \,, \qquad \text{for}  \qquad  r< r_{\rm crt} \,.
\end{equation}
Figure \ref{fig:tworegions} illustrates the relevant spacetime regions (and the corresponding `maximal' surfaces). Despite the negative contribution along this part of the hypersurface, the codimension-one observable defined in eq.~(\ref{eq:generlizedCV}) remains positive because $\mC_{\mt{gen}}$ is always dominated by the universal UV divergence near the conformal boundary. Further, it is clear that the integrand is always positive in the region near the asymptotic boundary.

We will show that linear growth of the observable at late times is prevented by the integrand becoming negative near and inside the black hole. Our definition of $\mC_{\rm{gen}}$, \ie 
\begin{equation}\label{eq:generlizedCV}
	\mC_{\rm{gen}}=\max_{\partial\Sigma=\Scft} \[  \frac{V_x }{\GN L} \int_\Sigma d\sigma\,\(\frac{r}{L}\)^{d-1}\!\sqrt{-f(r){\dot v}^2+2\dot v\,\dot r}\ a(r)  \]\,, 
\end{equation}
is still associated with the maximization process. Since the sign of $a(r)$ changes when we move from the conformal boundary to the black hole interior, it is more convenient to decompose the above maximization into two distinct regions  with 
\begin{equation}
	\begin{split}
		\mC_{\rm{gen}}=\frac{V_x }{\GN L}  \max_{\partial\Sigma=\Scft} &\left[ 2 \int^{r_\mt{max}}_{r_{\rm crt}} \!\! dr \,\(\frac{r}{L}\)^{d-1}\!\sqrt{-f(r) \( \frac{dt}{dr}\)^2+ \frac{1}{f(r)} }\,a(r)        \right. \\
		& \quad+  2 \int^{r_{\rm crt} }_{r_\mt{min}}
		\!\!d\sigma\,\(\frac{r}{L}\)^{d-1}\!\sqrt{-f(r){\dot v}^2+2\dot v\,\dot r}\ a(r) 
		\Bigg]  \,, \\ 
	\end{split} \label{Bigdog}
\end{equation} 
where we have assumed as usual that the maximal surface will be symmetric about the $t=0$ surface at the center of the Penrose diagram. Hence we only consider the right half of the surface and each integral comes with an extra factor of two. Further,
the volume integral of the outside region was rewritten in terms of $(t,r)$ coordinates because this portion of the surface with $r>r_{\rm crt}$ entirely outside the horizon (where $f(r)>0$), and we regulated the radial integration by stopping at some large $r=r_\mt{max}$.

Our approach will now be to first maximize the two integrals in eq.~\reef{Bigdog} separately. Implicitly this requires choosing a specific time $t=t_\mt{crt}$ at the critical surface $r=r_\mt{crt}$. That is, the outer integral is maximized with the boundary conditions
$(t,r)=(t_\mt{crt},r_\mt{crt})$ at the critical surface and $(t,r)= (t_\mt{R}=\tau/2,r_\mt{max})$ at the asymptotic boundary. Similarly, the inner integral must be maximized with $(t,r)=(t_\mt{crt},r_\mt{crt})$ at the critical surface which forms its outer boundary. However, as a final step, we must then maximize the sum of these results by varying over the position of the joint at $r=r_\mt{crt}$, \ie we must vary over all possible values of $t_\mt{crt}$ which lie within the WDW patch.

Turning first to the inner integral, it is evident that since $a(r)<0$, the integrand will be negative unless the surface becomes null (\eg with $\dot v=0$). Hence in the inner region $r< r_{\rm crt}$, the maximization procedure will always push the surface to be null as much as possible, in order to prevent any negative contributions.
That is, inside the critical radius, the maximal surface is always a null surface which yields a vanishing integral, \ie 
\begin{equation}
	\max   \int^{r_{\rm crt} }_{r_\mt{min}}
		\!\!d\sigma\,\(\frac{r}{L}\)^{d-1}\!\sqrt{-f(r){\dot v}^2+2\dot v\,\dot r}\ a(r)   =  0 \,. \label{Bigdog2}
\end{equation}
In figure \ref{fig:tworegions}, we illustrate this with null surfaces propagating to the past, \ie satisfying $\dot v= 2\,\dot r/f(r)$. However, in general, it could be any piecewise null surface that extends between the appropriate boundary points, \ie $(t,r)=(t_\mt{crt},r_\mt{crt})$ on the critical surfaces in the left and right exterior regions. Note that the maximal value \reef{Bigdog2} of the inner integral vanishes irrespective of the time $t_\mt{crt}$.
Hence the value of the observable will be determined entirely by the maximal value of the outer integral in eq.~\reef{Bigdog}. 

Turning to the outer integral, it is straightforward to see that if we choose the boundary times at the critical surface and the asymptotic boundary to be the same, \ie $t_\mt{crt}=t_\mt{R}$,  the locally extremal surface corresponds to a constant time slice with ${dt}/{dr}=0$. Now determining the profile of the extremal surfaces with other choices of $t_\mt{crt}$ is more involved but it is clear that the trajectory of surfaces will move in both time and radius, \ie ${dt}/{dr}\ne 0$. However, since $f(r)>0$ everywhere along the radial integration, it is clear that introducing ${dt}/{dr}\ne 0$ reduces the value of the corresponding integral. That is, maximizing the outer integral over different values of $t_\mt{crt}$ chooses $t_\mt{crt}=t_\mt{R}$. Hence the maximal result for the outer integral and the full observable in eq.~\reef{Bigdog} is a constant, \viz
\begin{equation}
		\mC_{\rm{gen}} =  \frac{V_x }{\GN L}  \int_{r_{\rm crt}}^{r_{\rm max}}\!\!dr\, \frac{a(r)}{\sqrt{f(r)}}  =\ \text{constant} \,. 
\end{equation} 

Physically, it is easy to understand why the late-time linear growth disappears in this situation. Different from the typical case where the generalized volume of the wormhole region grows linearly at late times, the negative part (\ie $a(r)<0$) inside the critical radius terminates the contribution coming from the growth of the wormhole to the integrand. Instead, the gravitational observable $\mC_{\rm{gen}}$ remains constant throughout the time evolution.

%%%%%%%%%%%%%%%%%%%%%%%%%%%%%%%%%%%%%%%%%%%%%%
%%%%%%%%%%%%%%%%%%%%%%%%%%%%%%%%%%%%%%%%%%%%%%
\section{Probing the singularity with CMC slices}
% !TEX root = ../singularity10.tex
%

\label{sec:CMC}

In this section, we use the flexibility afforded to us by ``complexity=anything" to begin investigating what properties can be extracted about the black hole singularity. As noted previously, in order to probe the geometry of the singularity, one must tune the parameters appearing in the gravitational observables to push the final slice at $r=r_f$ close to the singularity.  As an example, we focus here on a simple set of observables defined by local geometric functionals evaluated on time slices with constant mean curvature (CMC). These observables can be obtained from eq.~\reef{eq:O1} by setting $F_{2,\pm}$ and $G_2$ (\ie the functions used to extremize the boundary surfaces $\Sigma_\pm$) to be positive constants and with this choice, the extremal surfaces are both CMC slices. However, following the general procedure described in \cite{Belin:2022xmt}, we evaluate the observable with $F_{1,-}= G_1 = 0$. That is, we first  identify the CMC slice of interest as the future boundary of a codimension-zero region which extremizes a weighted sum of its spacetime-volume and the volume of its past and future boundaries, \viz 
\begin{equation}
\mathcal{C}_{\mt{CMC}}=\frac{1}{\GN L}\left[ \alpha_+\int_{\Sigma_+}d^d\sigma\sqrt{h}+  \alpha_- \int_{\Sigma_-}d^d\sigma \sqrt{h} +\frac{\alpha_{\mt{B}}}{L}\int_{\mathcal{M}}d^{d+1}x\sqrt{-g}\right],\label{eq:cmc}
\end{equation}
where $\alpha_{\pm}$ and $\alpha_{\mt{B}}$ are positive constants. This observable was introduced and examined in detail in~\cite{Belin:2022xmt}. However, here our observable will be constructed by using only the future CMC slice $\Sigma_+$. As noted above, the $\Sigma_-$ surface is also a CMC slice -- see details in \cite{Belin:2022xmt} -- but we discard this surface as part of our observable (\ie with $F_{1,-}=G_1=0$) in the following. The mean curvature of the future boundary can be expressed as~\cite{Belin:2022xmt}
\begin{equation}\label{eq:K}
K_{\Sigma_+}=-\frac{\alpha_{\mt{B}}}{\alpha_+ L}=-\frac{d}{L}\,\gamma
\quad {\rm where}\ \ \gamma\equiv\frac{\alpha_{\mt{B}}}{d\, \alpha_+}\,.
\end{equation}
Introducing the new parameter $\gamma$ will prove useful in the following analysis.
As we shall see, by varying the value of the mean curvature we control the distance between the CMC slice and the singularity at late times, allowing us to probe the  geometry near the singularity by examining the late-time growth of these CMC observables. 

%%%%%%%%%%%%%%%%%%%%%%%%%%%%%%%%%%%%%%%%
\subsection{AdS Schwarzschild black hole} \label{golly1}
%%%%%%%%%%%%%%%%%%%%%%%%%%%%%%%%%%%%%%%%

To begin, we will consider the AdS Schwarzschild black hole whose metric is given by eqs.~\reef{metricBH} and \reef{blacken} with general $k$. That is, the following analysis holds for spherical, planar and hyperbolic black holes.  For all of these cases, the CMC slice $\Sigma_{+}$ solves the variational problem with Lagrangian obtained from \reef{eq:Lagragianpm} by setting
\begin{equation}
a_{+}(r)=1\,,\qquad b(r)=\gamma\left(\frac{r}{L}\right)^{d}\,.
\end{equation}
The familiar maximal volume slice for the CV proposal can be obtained by setting $\gamma=0$, corresponding to vanishing extrinsic curvature in eq.~\reef{eq:K}. As discussed in section~\ref{sec:generalization}, the variational problem is equivalent to the classical mechanics problem of a particle in a potential, \viz
\begin{equation}
\dot{r}^2+\mathcal{U}(P_v,r)=0\,,\label{eq:particle}
\end{equation}
where the effective potential is given by
\begin{equation}\label{eq:potentials}
\mathcal{U}(P_v,r)=U_0(r)-\left(P_v + \gamma \left(\frac{r}{L}\right)^{d}\right)^2
\qquad{\rm with}\quad U_0(r)=-f(r)\left(\frac{r}{L}\right)^{2(d-1)}\,.
\end{equation}
Further, as in eq.~\reef{eq:moment}, the conserved momentum can be written as
\begin{equation}
P_v=\dot{r}-\dot{v}f(r)-\gamma \left(\frac{r}{L}\right)^{d} \,.
\end{equation} 
As long as $P_v$ is chosen to lie in a range such that the potential eq.~\reef{eq:potentials} has at least one root, there will be a nonvanishing value of $r=r_{\rm min}$, corresponding to the point closest to the singularity. The corresponding boundary time is then evaluated as 
\begin{equation}\label{eq:bdytime}
	\tau =-2 \int_{r_{ \min }}^{\infty} d r \frac{P_v+ \gamma \left(\frac{r}{L}\right)^{d}}{f(r) \sqrt{-\,\mathcal{U}(P_v,r)}} \,.
\end{equation}
It is straightforward to show that the late-time limit corresponds to tuning $P_v$ so that the potential has a degenerate root at $r=r_{f}$. As we approach late times, $r_{\rm min}$ will generally decrease and approach the final value $r_f$.

For the AdS Schwarzschild black hole, we can confirm that $r_f$ approaches the singularity in the limit of large extrinsic curvature, \ie $\gamma\gg 1$ in eq.~\reef{eq:K}. To see this, we recall that $r_f$ corresponds to a degenerate root of the effective potential, \ie  $\mathcal{U}(P_v,r)|_{r=r_f}=0=\partial_r\, \mathcal{U}(P_v,r)|_{r=r_f}$, which can be combined to yield\footnote{Solving for $\gamma$ and rewriting in terms of the extrinsic curvature $K=-\frac{d}{L}\gamma$, one finds that eq.~\reef{eq:approxfeq} is equivalent to the following exact result for the extrinsic curvature of a constant $r$ slice:
\begin{equation}
K|_{r=\text{const.}}=\frac{2(d-1) f(r)+r f'(r)}{2r\sqrt{-f(r)}},
\end{equation}
with the choice of $r=r_f$.
As we shall see, this is consistent with the result of eq.~\reef{eq:latettime}, which shows that at late times the CMC-slice hugs the constant radius surface $r=r_f$. Here we see that the extrinsic curvature of the CMC-slice and the $r=r_f$ surface match in the late time limit.}
\begin{equation}
	4 (d-1)^2 f^2(r_f)+r_f^2 f'^2(r_f)+4r_f f(r_f) \left( \frac{d^2}{L^2}r_f \gamma^2+(d-1)f'(r_f)\right)=0.
	\label{eq:findrf}
\end{equation}
Next we observe that $r_f$ can only be a root if $r_f < r_h$. Otherwise both terms in the potential \reef{eq:potentials} are negative. Furthermore, if the above expression is to hold for $\gamma\gg 1$, we must have either $r_f\to 0$ or $r_f\to r_h $, corresponding to $f(r_f)\to0$ or $f(r_f)\to -\infty $. Which of these two values we approach in the large mean curvature limit depends on whether we are considering the late- or early-time limit. To be more specific, recall that the CMC slice with large mean curvature approaches the future boundary of the WdW patch \cite{Belin:2022xmt}. It is then clear that the slice approaches the past horizon as we move the boundary time to the far past, while it approaches the future singularity for late boundary times.\footnote{Accordingly, if we flipped the sign of the mean curvature, the CMC slice would approach the past boundary of the WdW patch when $\gamma\gg 1$. In that case, the slice approaches the past singularity for boundary times in the far past, and the future horizon at late boundary times.}
We can then expand eq.~\reef{eq:findrf} near the singularity to find\footnote{For $k=0$, an explicit formula for $r_f$ in terms of $\gamma$ can be found in~\cite{Belin:2022xmt}.}
\begin{equation}\label{eq:limitvalues}
	r_f\simeq \(\frac{L^2\omega^{d-2}}{4\gamma^2}\)^{\frac{1}{d}}\qquad \text{with}\ \ \gamma\gg1\,.
\end{equation}

We now turn to study the late-time behaviour of various CMC observables. These observables are defined by choosing local functionals to integrate over the CMC slice  -- in our case corresponding to geometrical quantities like volume, extrinsic curvature and the square of the Weyl tensor. Note that these new functionals do not enter into any extremization procedure, and are simply evaluated on the previously defined surface $\Sigma_+$. In general, a CMC observable is defined by
\begin{equation}
\mathcal{C}^+=\frac{1}{\GN L}\int_{\Sigma^+}d^d \sigma \sqrt{h}\,a_1(r)=\frac{\Omega_{k,d-1}L^{d-2}}{\GN }\int_{\Sigma^+}d\sigma \left(\frac{r}{L}\right)^{d-1}\sqrt{-f(r)\dot{v}^2+2\dot{v}\dot{r}}\,a_1(r)\,,\label{eq:observable}
\end{equation}
where $a_1$ can be chosen to be any arbitrary scalar functional of the background metric, as well as the embedding function, \eg the extrinsic curvature. Note that because of the symmetry of the backgrounds in eq.~\reef{metricBH} which we study here, $a_1$ is only a function of the radial coordinate $r$. As before, $\Omega_{k,d-1}$ is the dimensionless volume of the transverse dimensions. For example,  it is the volume of the transverse unit sphere for $k=1$, while for $k=0$ it is the (regulated) volume of the transverse plane.

%\smr{I found some equations below do not have the proper length dimension (Have some extra $L$ or miss a factor $L$.). I have corrected those. But please double check. As a reminder, our conventions are $[\tau]=[t]=[r]=[\sigma]=[L], [C]=[O]=[P_v]=[L^0], [\GN]= [L^{d-1}]$.}

Using the gauge fixing condition~\reef{eq:gauge01} as well as the equation of motion \reef{eq:particle}, we can obtain 
\begin{equation}
\mathcal{C}^+=-\frac{2 \Omega_{k,d-1}L^{d-2}}{\GN }\int^{\infty}_{r_{\rm min}}\frac{dr}{f(r)\sqrt{-\mathcal{U}(P_v,r)}}\,U_0(r)\,a_1(r),\label{eq:observable2}
\end{equation} 
Accordingly, the growth rate can be expressed as
\begin{equation}
\frac{d \mathcal{C}^+}{d \tau}=\frac{\Omega_{k,d-1}L^{d-2}}{\GN } \left[\frac{2 U_0(r)a_1(r)}{f(r)\sqrt{-\mathcal{U}(P_v,r)}}\frac{d r_{\rm min}}{d \tau}\bigg\vert_{r=r_{\rm min}}+2 \frac{d P_v}{d\tau}\int^{\infty}_{r_{\rm min}}dr\, \frac{U_0(r)a_1(r)\left(P_v+\gamma \left(\frac{r}{L}\right)^d\right)}{f(r)\left(-\mathcal{U}(P_v,r)\right)^{\frac{3}{2}}}\right]\,.\label{eq:dO1}
\end{equation}
One needs to be careful with the above expression, as both terms are potentially divergent due to $\mathcal{U}(P_v,r_{\rm min})=0$. Nevertheless, a more careful derivation shows that the sum is finite. 
We can establish this by first differentiating eq.~\eqref{eq:bdytime}, which gives
\begin{equation}\label{eq:dt/drmin}
\frac{d r_{\rm min}}{d \tau}=\frac{ f(\rmin)\sqrt{-\mathcal{U}(P_v,\rmin)}}{2\left(P_v+\gamma \left(\frac{\rmin}{L}\right)^d\right)}  \left(1-\frac{d P_v}{d\tau}\int^{\infty}_{r_{\rm min}}dr\, \frac{2\,U_0(r)}{f(r)\left(-\mathcal{U}(P_v,r)\right)^{\frac{3}{2}}}\right)\,.
\end{equation}
Substituting the result back into eq.~\reef{eq:dO1} gives 
\beqa
\frac{d\mathcal{C}^+}{d\tau}=\frac{\Omega_{k,d-1}L^{d-2}}{\GN }&&\Bigg( \frac{ U_0(r_{\rm min})\,a_1(r_{\rm min})}{\left(P_v+\gamma \left(\frac{r_{\rm min}}{L}\right)^d\right)}+\frac{dP_v}{d\tau}\int^{\infty}_{r_{\rm min}}dr\, \frac{2\,U_0(r)}{f(r)\left(-\mathcal{U}(P_v,r)\right)^{\frac{3}{2}}}
\label{eq:finite-dO1}\\
&&\times\quad {\small \left[a_1(r)\left(P_v+\gamma \left(\frac{r}{L}\right)^d\right)-a_1(r_{\rm min})\left(P_v+\gamma \left(\frac{r_{\rm min}}{L}\right)^d\right)\right]}\Bigg)\,, 
\nonumber
\eeqa
where both terms are finite. Note that the numerator of the integrand goes to zero for $r\to r_{\rm min}$. The second term vanishes in the late-time limit, so we are left with 
\begin{equation}
\underset{\tau\to\infty}{\lim}\frac{d\mathcal{C}^+}{d\tau}=\frac{\Omega_{k,d-1}L^{d-2}}{\GN }\frac{U_0(r_{f})a_1(r_{f})}{\left(P_{\infty}+\gamma \left(\frac{r_{ f}}{L}\right)^d\right)}=\frac{\Omega_{k,d-1}L^{d-2}}{\GN }\sqrt{-f(r_f)}\(\frac{r_f}{L}\)^{d-1}a_1(r_f)\,, \label{eq:latettime}
\end{equation} 
where the second equality is obtained by using the fact that the potential~\reef{eq:potentials} vanishes at $r=r_f$. We see that the final expression is simply the volume measure on the final slice $r=r_f$ multiplied by the geometric factor $a_1(r_f)$. Thus, the late-time growth has an intuitive explanation in terms of the CMC slice spreading across the final time surface at a constant rate with respect to the boundary time. The segments of the CMC slice that extend out to the asymptotic boundary are not contributing to the above equation, as their volume approaches a constant value at late times.

Now we can use eq.~\reef{eq:latettime} to study the late-time growth rate of the CMC observables as we vary the mean curvature to allow the CMC slice approach the singularity. In particular, we take the limit $\gamma\gg 1$,
corresponding to large mean curvature -- see eq.~\eqref{eq:K}. Recall that in this limit, eq.~\reef{eq:limitvalues} indicates
$r_f\simeq \({L^2\omega^{d-2}}/{4\gamma^2}\)^{1/d}$, meaning that the volume element on the CMC slice goes as
\begin{equation}
 \sqrt{-f(r_f)}\left(\frac{r_f}{L}\right)^{d-1}\simeq \frac{1}{2}\(\frac{\omega}{ L}\)^{d-2}\,\frac{1}{\gamma}\quad \text{for}\,\,\,\, \gamma\gg 1\,.
\end{equation}
Substituting this expression into eq.~\reef{eq:latettime} yields
\begin{equation}
\frac{d\mathcal{C}^+}{d\tau}\simeq\frac{\Omega_{k,d-1}L^{d-2}}{\GN }\, \frac{1}{2}\(\frac{\omega}{ L}\)^{d-2}\frac{1}{\gamma}\ a_1\!\(\(\frac{L^2\omega^{d-2}}{4\gamma^2}\)^{1/d}\) =  \frac{8\pi M }{(d-1)}\,\frac{a_1(r_f)}{\gamma}\,. 
\label{eq:limitdO}
\end{equation}

Now we consider three explicit examples of geometric observables on the CMC slice $\Sigma^+$. First, we can obtain the volume of the CMC slice by choosing $a_1(r)=1$ . Using eq.~\reef{eq:limitdO}, we find
\begin{equation}
a_1=1\ \ :\quad\frac{d\mathcal{C}^+}{d\tau}\simeq\frac{8\pi M }{(d-1)}\,\frac{1}{\gamma}\to 0\,. \label{volly}
\end{equation}
This is not surprising, as in this limit is the CMC slice approaches the future boundary of the WdW patch, which consists of two null segments and the segment hugging the singularity, where the transverse sphere has vanishing volume. 

Secondly, we examine the extrinsic curvature by taking $a_1=-LK$, where we chose the minus sign to make the late-time growth positive. Using eq.~\reef{eq:K}, we find $a_1(r)=d\,\gamma$, which then gives a finite result for the late-time growth, namely
\begin{equation} 
a_1=-L K\ \ :\quad\frac{d\mathcal{C}^+}{d\tau} \simeq\frac{8\pi M\,d}{(d-1)}\,.
\label{Kobs}
\end{equation}
Hence the late-time growth is linear as expected, even in this $\gamma\gg 1$ limit. The finiteness arises because, as described above, the late-time growth is controlled by the portion of the extremal surface spreading out along the singularity. Here the divergence in $K$ precisely matches the vanishing volume element near the singularity. For more discussion about the finiteness of $K\sqrt{h}$ term on a generic singularity, see appendix \ref{sec:nearsingularity}. 

Finally, we consider the Weyl-squared term $C^2= C_{\mu\nu\rho\sigma}\,C^{\mu\nu\rho\sigma}$. With $a_1(r)=L^4 C^2=d(d-1)^2(d-2)\frac{L^4\omega^{2(d-2)}}{r^{2d}}$, we find the following late-time growth
\begin{equation} 
a_1=L^4C^2\ \ :\quad\frac{d\mathcal{C}^+}{d\tau}\simeq d(d-1)(d-2)\,128\pi M  \gamma^3 \to\infty\,. \label{Ctwo0}
\end{equation}
Generally, we expect a similarly divergent result if we construct the observable $\mathcal{C}^+$ with higher products involving $n$ factors of the Weyl tensor. These will diverge as $1/r^{n d}$ near the singularity, meaning that the corresponding CMC observable will grow as $\gamma^{2 n -1}$ with $\gamma\gg1$. In summary, we see that the different CMC observables show a wide range of behaviours as the CMC slice is pushed closer to the singularity: either vanishing, approaching a constant value, or growing without bound (\ie diverging). Further, however, the growth with $\gamma$ in the latter case is characteristic of the geometry at the curvature singularity -- see section \ref{sec:disc} for further discussion.

%%%%%%%%%%%%%%%%%%%%%%%%%%%%%%%%%%%%%%%%
\subsection{AdS Reissner-Nordstr\"{o}m black hole} \label{golly2}
%%%%%%%%%%%%%%%%%%%%%%%%%%%%%%%%%%%%%%%%
Having studied the behaviour of various CMC observables near the singularity of the AdS Schwarzschild black hole, we now carry out a similar analysis for the AdS Reissner-Nordstr\"{o}m (RN) black hole, whose metric is again given by eq.~\reef{metricBH} but now with the blackening factor 
\beq
f_\Q(r)= k+ \frac{r^2}{L^2}- \frac{\omega^{d-2}}{r^{d-2}} + \frac{q^2}{r^{2(d-2)}}\,. \label{charged8}
\eeq
Again the horizon geometry is either spherical, planar or hyperbolic with $k=+1,$ 0 or --1, respectively. This geometry is a solution to Einstein gravity with a negative cosmological constant and a $U(1)$ gauge field. The bulk action is given by 
\begin{equation}
	I=I_\mt{grav}- \frac1{16\pi\GN}\int d^{d+1}x\,\sqrt{-g}\,F_{ab}F^{ab}\,,
\end{equation}
with $F_{ab}=\partial_a A_b-\partial_b A_a$ as usual. The gauge potential in the AdS RN background can be written as (\eg see \cite{Chamblin:1999tk}),
\begin{equation}
A_t=\sqrt{\frac{d-1}{2(d-2)}}\left(\frac{1}{r_+^{d-2}}-\frac{1}{r^{d-2}}\right)q\,,
\end{equation}
where $r_+$ is the outer horizon radius (defined below). In the dual description, this gauge field introduces a chemical potential, which is given by the `non-normalizable' mode, \ie $\mu=\lim_{r\to\infty} A_t$. Accordingly, boundary state dual to the AdS RN black hole is the so-called charged thermofield double state, where the sum over states is weighted by not only their energy but also their $U(1)$ charge \cite{Carmi:2017jqz}, \viz
\begin{equation}
\ket{\mathrm{cTFD}(t_{\mt{L}},t_{\mt{R}})}=\frac{1}{\sqrt{Z}}\sum_{\alpha,\sigma}e^{-\beta(E_\alpha - \mu Q_\sigma)/2}e^{-i E_\alpha (t_{\mt{L}}+t_{\mt{R}})}\ket{E_\alpha,-Q_\sigma}_{\mt{L}}\ket{E_\alpha,Q_\sigma}_{\mt{R}},
\end{equation}
where the subscripts $\mathrm{L}$ and $\mathrm{R}$ label quantities associated with the left and right boundaries, respectively.  If we trace out one of the boundary Hilbert spaces, we are left with a density matrix describing a grand canonical ensemble. 

In what follows, we will assume the RN black hole is non-extremal. Then, in contrast to the Schwarzschild geometry, the RN black hole has a timelike singularity, as well as inner and outer horizons at $r=r_{\pm}$ where $f(r_\pm)=0$. As a result, the singularity is  inaccessible to the CMC slice surfaces, since they are anchored to the asymptotic boundaries and remain spacelike (or null) throughout the bulk. Indeed, the CMC slices can only probe the black hole interior down to the the inner horizon $r=r_-$, which one again reaches by considering the limit of large mean curvature (\ie $\gamma\gg1$). In contrast to the AdS Schwarzschild case, we can expect all of the CMC observables to  be well behaved (\ie finite) in this limit, as the geometry of the AdS RN black hole remains nonsingular near the inner horizon. 

That the CMC slices cannot probe beyond the inner horizon is reflected in the fact that the solutions for the turning point equation, \ie 
\begin{equation}
\mathcal{U}_\Q(P,r_\mt{min})\equiv-\left(\frac{r_\mt{min}}{L}\right)^{2(d-1)}f_\Q(r_\mt{min})-\left(P+\gamma \left(\frac{r_\mt{min}}{L}\right)^{d} \right)^2=0,
\label{eq:Prmin}
\end{equation}
only occur for $r_-<r_\mt{min}<r_+$ where the factor $f_\Q(r_\mt{min})$ is negative. Otherwise, both terms in eq.~\reef{eq:Prmin} are strictly negative beyond this range, and hence there are no solutions inside of the inner horizon, \ie with $r_\mt{min}<r_-$. 

Furthermore, we can confirm that $r_f$ approaches $r_-$ in the limit of large extrinsic curvature (\ie $\gamma\gg1$) by an analogous argument to the AdS Schwarzschild case previously. That is, we enforce the condition ${\partial_r\, \mathcal{U}_\Q(P,r)|_{r=r_f}=0}$, which can be combined with $r_\mt{min}=r_f$ in eq.~\reef{eq:Prmin} to yield eq.~\reef{eq:findrf} with the appropriate substitution $f(r)\to f_\Q(r)$. We then observe that for this expression to hold with $\gamma\gg1$, we must have $f_\Q(r_f)\to0$. Hence in this limit, we must be approaching either the inner or outer horizon $r=r_{\pm}$ as they correspond to the two roots of $f_\Q$. Which root we approach in the large mean curvature limit depends on whether we are taking the late or early time limit, analogously to the case for the Schwarzschild black hole. In our case, we are interested in the late time limit for which $r_f\to r_-$. Further, we find the following relation in the  regime of large mean curvature
\begin{equation}
f_\Q'^2(r_f)+4\frac{d^2}{L^2}f_\Q(r_f)\gamma^2\simeq0,\qquad \text{for}\quad \gamma\gg 1\,.
\label{eq:approxfeq}
\end{equation}
We can expand the above equation around the inner horizon (\ie $r_f\sim r_-$) to find 
\begin{equation}\label{inner1}
	r_f\simeq r_--\frac{L^2}{4d^2\,\gamma^2}\,f_\Q'(r_-)\quad \text{with}\,\,\gamma\gg 1\,.
\end{equation}
Note that $f_\Q'(r_-)<0$ and hence $r_f$ is slightly larger than $r_-$, as expected. The above result is all we need to evaluate the late-time growth of CMC observables in the RN-AdS geometry. In particular, we can utilize eq.~\reef{eq:latettime}, again with the appropriate substitution of $f(r)\to f_\Q(r)$. As in the Schwarzschild case, the volume element on the CMC slice approaches zero for $\gamma\gg1$,\viz
\begin{equation}
 \sqrt{-f_\Q(r_f)}\left(\frac{r_f}{L}\right)^{d-1}\simeq |f'_\Q(r_-)|\left(\frac{r_-}{L}\right)^{d-1}\frac{L}{2 d \,\gamma}\,.
\end{equation}
However, in that case, it was the volume element on surfaces parallel to the singularity which went to zero. Here, the volume element vanishes because we are approaching a null surface, \ie the inner horizon. Note that the late time volume element decays here with the same power of $\gamma$ as in the uncharged case. Hence we find the following expression for the late-time growth: 
\begin{equation}
\frac{d\mathcal{C}^\Q}{d\tau}\simeq  \frac{\Omega_{k,d-1}L^{d-2}}{\GN } |f'_\Q(r_-)|\left(\frac{r_-}{L}\right)^{d-1}\frac{L}{2d \gamma} a_1\(r_-\) =\frac{8\pi}{d}\,S_-T_-\,  \frac{a_1(r_-)}\gamma\,.
\label{eq:genlatetimeRN}
\end{equation}
Here we have written the final expression in terms of the Bekenstein-Hawking entropy and Hawking temperature associated with the inner horizon, \ie
\begin{equation}\label{innerST}
	S_- = \frac{\Omega_{k,d-1}\,r_-^{d-1}}{4\,\GN }\qquad{\rm and}\qquad
	T_- = \frac{|f'_\Q(r_-)|}{4\pi}\,.
\end{equation}

Now we examine the behaviour of the same three observables considered above for the AdS Schwarzschild case. For the volume of the CMC slice, we find
\begin{equation}
a_1=1 :\quad\frac{d\mathcal{C}^\Q}{d\tau}\simeq \frac{8\pi\,S_-T_-}{d}\, \frac{1}\gamma \to 0\,, 
\label{eq:Qvol}
\end{equation}
by setting $a_1(r)=1$. Again, this vanishing is expected since as shown in eq.~\reef{inner1}, for $\gamma\gg1$, $r_f$ approaches the inner horizon which is a null surface.  
For the extrinsic curvature observable, we have $a_1(r)=-L K=d \gamma$ and so we find
\begin{equation} 
a_1=-L K :\quad\frac{d\mathcal{C}^\Q}{d\tau}\simeq  8\pi S_-T_-\,. \label{golly789}
\end{equation}
That is, as for the uncharged black holes in eq.~\reef{Kobs}, we again find the late-time growth rate is finite for this observable, however, with a different coefficient.
Finally, we consider the square of the Weyl tensor, which gives 
\begin{equation}\label{CC22}
a_1(r)=L^4C^2=d(d-1)^2(d-2)\,\frac{L^4\omega^{2(d-2)}}{r^{2d}}\,\left(1-\frac{2(2d-3)}{d}\,\frac{q^2}{(\omega r)^{d-2}}\right)^2\,,
\end{equation}
for the AdS RN geometry. We then find the late-time growth rate to be
\begin{equation}
a_1=L^4 C^2: \quad  \frac{d\mathcal{C}^\Q}{d\tau}\simeq 8\pi (d-1)^2(d-2)\,\frac{L^4 \omega^{2(d-2)}}{r_-^{2d}}\left(1-\frac{2(2d-3)q^2}{d\, (\omega r_-)^{d-2}}\right)^2\,\frac{S_-T_-}{\gamma}\to 0\,. \label{golly4}
\end{equation}
This vanishing of the growth rate is again expected since the CMC slice is approaching the inner horizon where the (square of the) Weyl tensor remains finite while the volume element vanishes. Hence we simply see the same scaling with $\gamma$ here as for the volume in eq.~\reef{eq:Qvol}. However, the coefficient here reveals the value of $C^2$ at the inner horizon.

The above behaviour (\ie the $1/\gamma$ decay in eq.~\reef{eq:Qvol}) will also arise for observables involving higher powers of the Weyl tensor. Indeed, one does not expect to be able to construct an observable from the background curvatures alone that leads to a divergent growth rate in this case.  Of course, observables involving higher powers of the extrinsic curvature (\eg $a_1(r)=L^2\,K^2$) will yield a divergent growth rate. This solution can also be probed in an interesting way by  observables constructed with scalar functions involving the matter fields. For example, $a_1(r)= -L^2\, F_{ab} F^{ab}= \frac{(d-1)(d-2)L^2q^2}{2\,r^{2(d-1)}}$ yields
\begin{equation}
a_1=-L^2\, F_{ab} F^{ab}\ \ :\quad\frac{d\mathcal{C}^\Q}{d\tau}\simeq \frac{4\pi (d-1)(d-2)L^2q^2}{d\,r_-^{2(d-1)}}\, \frac{S_-T_-}\gamma \to 0\,,
\label{gollyF2}
\end{equation}
with the same the $1/\gamma$ decay expected from eq.~\reef{eq:Qvol}.
Of course, these matter observables provide diagnostics which distinguish the interior of the AdS RN black hole or other nonvacuum solutions.

%%%%%%%%%%%%%%%%%%%%%%%%%%%%%%%%%%%%%%%%%%%%%%

%%%%%%%%%%%%%%%%%%%%%%%%%%%%%%%%%%%%%%%%%%%%%%	
\section{Discussion}
% !TEX root = ../singularity10.tex
%
\label{sec:disc}

Our paper aimed to better understand how the complexity=anything approach
proposed in \cite{Belin:2021bga,Belin:2022xmt} can be used to examine the interior geometry of asymptotically AdS black holes, particularly their spacetime singularities.  We might contrast this new approach with the behaviour of the previously known holography complexity conjectures for, \eg the AdS Schwarzschild black hole given by eqs.~\reef{eq:AdSBH} and \reef{blacken}. Recall that with the CV proposal, the maximal volume surface does not approach very close to the singularity, \eg $r_f=r_h/2^{1/d}$ for the planar case (\ie $k=0$). Rather the extremal surface prefers to stay away from the (spacelike) singularity where the volume measure shrinks to zero. On the other hand, the WDW patch, appearing in both the CA and CV2.0 proposals, intersects with the singularity by the definition of this spacetime region. However, neither approach offers any specialized insights into the nature of the singularity. Our investigations here provide an initial demonstration that the flexibility of the complexity=anything proposal allows one to extract information about the characters of black hole singularities.

As reviewed in section \ref{recap}, the linear growth observed in both codimension-zero and codimension-one observables at late times is due to the linear expansion of the wormhole region of extremal surfaces. Our discussion of the linear growth is somewhat more general than in \cite{Belin:2021bga,Belin:2022xmt} because we did not choose a specific blackening factor $f(r)$ in the metric \reef{metricBH}. The only implicit assumption is that there is an `interior' region where $f(r)<0$ so that the effective potential \reef{potent} can have a positive maximum in this region. However, for generic gravitational observables, this linear growth offers limited insight into the black hole interior geometry because this maximum (which defines the final slice) generally acts as a barrier to accessing the geometry near the singularity. Hence this generic behaviour is not very different from that described above for the CV proposal.

\subsection*{Shortcomings with previous analysis}

We found that the analysis presented in \cite{Belin:2021bga,Belin:2022xmt} is somewhat incomplete. Our attention was drawn to this point in section \ref{sec:end}, where we turned to a puzzle which first appeared in \cite{Belin:2021bga}. There it was found that the codimension-one observable in eq.~\reef{eq:generalziedC2} only yielded extremal surfaces at late times with a limited range \reef{range1} of the coupling defining the strength of the $C^2$ term. Interestingly, as shown in figure \ref{fig:Pvtau}, if the coupling is tuned to be slightly outside of the allowed range, the complexity appears to grow linearly for a finite time. After a certain critical time $\tau_{\rm max}$, there is no corresponding extremal surface and hence the standard analysis introduced in \cite{Belin:2021bga,Belin:2022xmt} yields no result for the growth rate. Further, as we extend $\tlam$ beyond the allowed range, $\tau_{\rm max}$ rapidly decreases and the phase of linear growth disappears.

However, a more careful examination reveals that the surfaces yielding the maximal value of the observable are pushed to the edge of the allowed phase space. Hence, these `maximal' surfaces are no longer locally extremal. That is, they do not solve the equations derived from extremizing the observable, as described in section \ref{recap}. In the planar AdS-Schwarzschild background, for $\tlam>0$, the maximal surfaces are pushed to the future boundary of the corresponding WDW patch. Hence they are null sheets falling from the asymptotic boundary to the black hole singularity and then the central component hugs the singularity between these two -- see figure \ref{fig:tall1}. Unfortunately, the observable and the growth rate diverge when evaluated on these maximal surfaces, but this behaviour can be regulated by adding a term involving a higher power of the curvature tensor, as described in section \ref{regular1}. We emphasize that this behaviour arises for any positive value of $\tlam$, not just for $\tlam>\tlam_{\mt{crt}1}$, and for all times (where the WDW patch reaches the singularity), not only beyond some $\tau_\mt{max}$.

The case of $\tlam<-1$ was even more interesting, as discussed in section \ref{sec:negative}. In this case, the integrand of the observable becomes negative within a certain radius $r_\mt{crt}$ which lies outside of the horizon. The maximization procedure then includes three steps: For $r>r_\mt{crt}$, we solve the standard extremization equations for a given boundary time and a fixed time on the critical surface $r=r_\mt{crt}$. For $r<r_\mt{crt}$, the maximal value is found by choosing the surface to be (piecewise) null so that the net contribution from this region is zero. In particular, the latter vanishing result can be achieved independently of the `boundary' times at $r=r_\mt{crt}$. Finally extremizing over the time on the critical surface, we found that the exterior solution is chosen to be a constant $t$ surface, which maximizes the contribution from this region. As a result, the observable is constant in this regime and the growth rate vanishes for $\tlam<-1$. 

While we illustrated this behaviour with a particular example of a codimension-one observable in section \ref{sec:end}, the situation can occur quite generally with both codimension-one and codimension-zero observables.  In particular, in certain circumstances, the surfaces yielding the maximal value for the observable are no longer locally extremal. That is, the `maximal' surfaces do not solve the equations derived from extremizing the observable, as described in section \ref{recap}. Our example shows that the analysis there may fail in situations where $a(r)$ diverges (positively) in approaching the singularity or where $a(r)$ becomes negative outside of the horizon.

Furthermore, let us note that the results described above are dependent on the choice of background. For example, for a non-extremal charged black hole, as in eq.~\reef{charged8}, it is straightforward to show that the growth rate remains finite for $\tlam>0$. The essential point is that, even though we have $a(r\to0)\to+\infty$ at the (timelike) singularity, the singularity is shielded by the inner horizon and this region is not accessed by spacelike surfaces connected to the asymptotic boundaries. Hence, these observables with $\tlam>0$ may still be used as a probe to distinguish different black hole interiors. That is, the growth rate is divergent with a spacelike singularity behind the event horizon where $C^2$ diverges, while it remains finite when the singularity is hidden by an inner horizon which the spacelike surfaces will not penetrate.\footnote{It is expected that general perturbations of the background will cause the Cauchy horizon to become singular \cite{Poisson:1989zz,Poisson:1990eh,Ori:1991zz,Brady:1995ni}. It would be interesting to investigate how the observable \reef{eq:generalziedC2} behaves in this situation. \label{footy99}} Let us add that for $\tlam<-1$, the growth rate of the complexity remains zero for the charged black holes.

However, one must ask whether or not the above results make sense from the perspective of complexity in the boundary theory. It seems that the observables with $\tlam<-1$ simply fail to be viable candidates for the holographic dual of boundary complexity by the standard criteria considered in \cite{Belin:2021bga,Belin:2022xmt}, \ie they do not exhibit linear growth at late times. The case of 
$\tlam>0$ is perhaps more interesting but the interpretation remains unclear. Here the late-time growth rate diverges for the usual thermofield double state \reef{eq:TFD}, but the rate remains finite when a chemical potential is added. One might imagine that this behaviour arises with a particular choice of the microscopic gates used to construct the target state. That is, certain key gates always push the underlying circuits towards preparing entangled states where the chemical potential is turned on. Nonetheless, since there must be gates available to construct states with either a positive or negative chemical potential, it is not clear why some combination of these would not yield states with zero chemical potential. Perhaps the divergent complexity reflects an excessive fine-tuning required to achieve a vanishing chemical potential in this situation.

\subsection*{Probing the singularity}

To probe the black hole singularities in section \ref{sec:CMC}, we considered constant mean curvature (CMC) surfaces and a limiting procedure which brought the final surface arbitrarily close to the singularity (in the case of the AdS Schwarzschild black hole). Our construction used the simplest codimension-zero observables \eqref{eq:cmc} (introduced in \cite{Belin:2022xmt}) to determine the extremal surfaces, for which both the future and past boundaries are CMC slices. By fine-tuning the parameter associated with the future boundary (\ie taking $\alpha_+\ll1$ or $\gamma \gg1$), one can bring the CMC slice close to the future/past light cone, which reaches the singularity in the AdS-Schwarzschild black hole background. In this way, a large portion of the resulting CMC slice hugs the spacelike singularity at late times. To probe this geometry, we examined the growth rate of the observable defined by evaluating various curvature scalars on this surface -- see section \ref{golly1}. We found that the late-time growth rate can either become vanishingly small, converge to a finite constant or grow arbitrarily large (\eg for $a \sim 1,$ $K$ or $C^2$, respectively). Further, the decay/divergence rate can be parameterized in terms of the dimensionless parameter $\gamma$ and encodes information about the spacetime geometry in the vicinity of the singularity. For example, the power $1/\gamma$ in eq.~\eqref{volly} indicates that the volume measure on constant radius surfaces decays as $r^{d/2}$ near the singularity. Combined with eq.~\eqref{Ctwo0} where the growth rate diverges at $\gamma^3$, we see that $C^2$ diverges as $1/r^{2d}$ in approaching the singularity.

These results may be contrasted with those for the AdS Reissner-Nordstr\"{o}m background in section \ref{golly2}. For these (nonextremal) charged black holes the timelike singularity is shielded by an inner horizon which the CMC slices will not penetrate. Hence with $\gamma\gg1$, the CMC surfaces again approach the future lightcone but hug the inner horizon rather than probing the singularity. In this case, we see in eq.~\reef{eq:Qvol} that for the observable with $a(r)=1$, the late-time growth rate decays as $1/\gamma$ (precisely as above), which reflects the vanishing of the volume measure as the CMC slice approaches a null surface. Further, with $a(r)=L^4C^2$ in eq.~\reef{golly4}, the growth rate still decays as $1/\gamma$ which reflects the fact that the curvature remains finite in the vicinity of the null horizon. Hence these observables are demonstrating that the interior geometry of these charged black holes is very different from that in the uncharged case.

In any event, rather than viewing the ambiguities in defining holographic complexity as a shortcoming, our approach here views this ambiguity as a feature on which we can capitalize. Focusing on a single gravitational observable would not yield much information about the black hole interior. Rather it is only by comparing the growth rate for different gravitational observables that we are able to extract a detailed picture of the interior geometry. Of course, we are not comparing completely distinct observables. Instead, we are making use of the complexity=anything approach to evaluate different geometric features of the same extremal surfaces. All of these observables are equally viable candidates for the holographic dual of the quantum complexity of the corresponding boundary states. Hence it would be interesting to understand which parameters of the underlying complexity model in the boundary theory are changed when we choose different geometric functionals for the gravitational observables. To study this question, it is probably best to combine the various functionals considered in eqs.~\reef{volly}, \reef{Kobs} and \reef{Ctwo0} together as
\begin{equation}\label{unify}
	a_1= \alpha_1 -\alpha_2\,L\,K +\alpha_3\,L^4\,C^2 +\cdots\,,
\end{equation}
where $\alpha_i$ are (dimensionless) positive parameters which may smoothly vary. We have included the ellipsis to indicate that one may wish to further extend this functional by including other geometric terms, \eg higher powers of the Weyl tensor.

Let us comment that we can emulate the above approach using the idea of `regulated' codimension-one observables, introduced in section \ref{regular1}. As discussed above, the maximal surfaces for the observable in eq.~\reef{eq:generalziedC2} with $\tlam>0$ were pushed into the singularity of the AdS Schwarzschild black hole.
However, this behaviour could be regulated by introducing an extra $C^4$ term with a small coupling, as in eq.~\reef{observe4}. In this case, the complexity remains finite but tuning of the new coupling allows the radius of the final surface to come arbitrarily close to the singularity, with $r_f\sim \tlam_4^{\,1/2d} r_h$ as shown in eq.~\reef{peak}. First, let us note that this procedure is not unique. It is straightforward to show that if the `regularizing' term in eq.~\reef{observe4} is replaced by $C^{2n} \equiv (C_{\mu\nu\alpha\beta} C^{\mu\nu\alpha\beta} )^n$, then in the regime $0<\tlam_{2n}\ll1$ (and $\tlam>0$), the global maximum in the effective potential appears at 
\begin{equation}\label{peakN}
w_f^{2(n-1)} \simeq \frac{(4n-1)}{3}\,\frac{\tlam_{2n}}{\tlam}\qquad \longrightarrow
\quad r_f \simeq \left(\frac{(4n-1)}{3}\,\frac{\tlam_{2n}}{\tlam}\right)^{\frac1{2(n-1)d}}\,r_h\,,
\end{equation}
and further, the late-time growth rate is given by
\begin{equation}\label{laterateN}
	\lim_{\tau \to \infty}  \(    \frac{d	\mC_{\rm gen}}{d\tau} \)  = \frac{64\pi}{(d-1)}\,\frac{n-1}{4n-1}\,\left(\frac{3}{(4n-1)}\,\frac{\tlam}{\tlam_{2n}}\right)^{\frac3{4(n-1)}}\,\tlam\,M \,.
\end{equation}
Hence for these generalized observables, we have that as $\tlam_{2n}\to0$, $r_f\sim \tlam_{2n}^{\ 1/2(n-1)d}\to 0$ and the time rate of change diverges as $\tlam_{2n}^{\ -3/4(n-1)}$. 

A more careful analysis, comparing these results for different regulators (\ie different values of $n$) may allow one to extract information about the geometry in the vicinity of the singularity. However, a simpler approach is to emulate the discussion in section \ref{sec:CMC}. That is, for a fixed regulator, we examine different observables by evaluating different curvature scalars at the extremal surface. For example with eq.~\reef{observe4}, we find that the late-time growth rate 
\begin{equation}\label{golly5}
	\lim_{\tau \to \infty}  \(    \frac{d	\mC_{\rm gen}}{d\tau} \)  \sim
	\tlam_4^{\,\frac14}\,,\quad 1\quad{\rm and}\quad \frac{1}{\tlam_4^{\,\frac34}}\,,
\end{equation}
for $a(r)=1$, $-L\,K$ and $L^4C^2$, respectively. 
We might note that there is a close correspondence between the powers of $\tlam_4$ above and the powers of $\gamma$ appearing in eqs.~\reef{volly}, \reef{Kobs} and \reef{Ctwo0}. In section \ref{golly1}, we found $r_f\sim\gamma^{-2/d}$ while here we have $r_f\sim\tlam_4^{\,1/2d}$, and hence the corresponding powers differ by a factor of $-1/4$. Of course, this approach allows us to extract the same information as above about the geometry near the singularity.

\subsection*{Anisotropic singularities}

All of the black holes \reef{metricBH} considered in our paper are characterized by a high degree of symmetry, which constrains the geometry near the singularity. Although these singularities are not completely isotropic (\ie the $t$ direction is distinguished from the rest of the spatial directions), it would be interesting to probe more generic singularities using complexity=anything. For solutions of the Einstein field equations, it is conjectured that the most generic spacelike singularities take the form of BKL (Belinski-Khalatnikov-Lifshitz) singularities \cite{Lifshitz:1963ps,Belinsky:1970ew,Belinsky:1982pk}. The BKL conjecture states that these generic spacelike singularities possess three properties, \eg see reviews in \cite{Montani:2007vu,Henneaux:2007ej,Belinski:2017fas}. Approaching the singularity, the physics is 1) ultralocal (\ie the evolution of each spatial point is governed by a system of ordinary differential equations with respect to time), 2) chaotically oscillatory (\ie generically at each point, the asymptotic behaviour is a chaotic, infinite, oscillatory succession of Kasner epochs), and 3) the evolution is dominated by the vacuum equations (\ie the matter contributions can be neglected asymptotically). However, for simplicity, we will restrict our comments to a simpler class of geometries known as Kasner solutions.

The Kasner geometry describes the most general anisotropic but homogeneous metric near a cosmological singularity, \ie 
\begin{equation}\label{eq:Kasner}
	ds^2 =  -d\tau^2 + \sum_{i=1}^d \tau^{2p_i} dx_i^2   \,, 
\end{equation}
where the cosmological singularity is located at the spacelike hypersurface $\tau=0$, and the constants $p_i$ are referred to as the Kasner indices. Demanding that the Kasner metric \eqref{eq:Kasner} is a solution of the vacuum Einstein equations with a {\it vanishing} cosmological constant\footnote{Including matter terms,  the second constraint is generally not satisfied, \ie $\sum_{i=1}^d  p_i^2\ne 1$  \cite{Belinski:2017fas}.} imposes the constraints: 
\begin{equation}\label{holly9}
	\sum_{i=1}^d  p_i=1=\sum_{i=1}^d  p_i^2\,.
\end{equation}
We note that with a nonvanishing cosmological constant (as typically arises in a holographic setting), the Kasner metric \eqref{eq:Kasner} remains an asymptotic solution near the singularity. Expanding Einstein's equations in inverse powers of $\tau$, this metric still captures the leading and subleading behavior near the singularity, with the cosmological constant only appearing at the next order.\footnote{An {\it exact} solution with a negative cosmological constant incorporating Kasner-like behaviour is 
\begin{equation}
ds^2 =  \frac{1}{z^2} \(   -d\tau^2 + \sum_{i=1}^d \tau^{2p_i} dx_i^2   + d z^2 \)  \,.
\end{equation}
This solution is studied in a holographic context by \cite{Engelhardt:2014mea,Engelhardt:2015gta}. } 
The Kasner-type singularity has also been investigated in the context of the AdS/CFT correspondence, \eg \cite{Das:2006dz,Craps:2007ch,Awad:2008jf,Engelhardt:2014mea,Engelhardt:2015gta,Barbon:2015ria,Shaghoulian:2016umj,Frenkel:2020ysx,Caputa:2021pad}. 

For the present purposes, it is interesting to ask if the black hole singularity took the form of a Kasner singularity, how we could extract information about this geometry using complexity=anything? For example, would we be able to determine the indices $p_i$ associated with distinct spatial directions? Of course, the anisotropic nature of the background would make finding extremal surfaces a challenging task. However, we observe that, up to this point, we have only considered surfaces that are anchored to a constant time slice on the boundary. Even if the singularity were anisotropic, the corresponding observables would only yield information that is averaged over the different directions. However, we could extend our present analysis further to produce anisotropic probes of the bulk by anchoring the extremal surfaces to different boundary Cauchy surfaces that are, \eg tilted along different spatial directions. That is, instead of choosing $t=$ constant, we would choose $t = f(x^i)$. The corresponding extremal surfaces would then be anisotropic in the bulk as well. Although it would still be a challenging task, it should be possible to extract information about the anisotropic nature of the black hole interior and the singularity with these new probes of the bulk geometry. Rotating black holes could provide an interesting setting in which to develop our understanding of such anisotropic extremal surfaces.

To close here, let us note that as emphasized in section \ref{sec:CMC}, the complexity=anything proposal establishes a two-step procedure that decouples the geometric quantity defining the extremal surface from the geometric quantity evaluated on the surface. In this spirit, it is interesting to observe that choosing $a_1=-L\,K$ yields a finite growth rate even as the extremal surface approaches the singularity in section \ref{golly1}. A similar finite result would be produced in section \ref{regular1} in the limit $\tlam_4\to0$, if the observable was chosen with $a_1=-L\,K$ on the extremal surface. Of course, the same observation was already made for the finiteness of the Gibbons-Hawking-York boundary term of the future boundary of the WDW patch for complexity=action, \eg see \cite{Brown:2015lvg}.\footnote{For further investigations of how the complexity=action proposal detects and probes black hole singularities, see also \cite{Barbon:2018mxk,Bolognesi:2018ion,Caceres:2022smh,An:2022lvo,Katoch:2023dfh}.}  While the (trace of the) extrinsic curvature diverges as the corresponding surface approaches the spacelike singularity, this divergence is precisely balanced by the vanishing of the volume measure. In fact, this finiteness can be related to the first constraint in eq.~\reef{holly9} for a Kasner singularity, \ie $\sum p_i=1$, as we discuss in appendix \ref{sec:nearsingularity}. Generally, we show there that the finiteness of $K\sqrt{h}$ requires that the matter contributions do not diverge too quickly near the singularity. 

This discussion also suggests that it may be interesting to study the codimension-one observable of the form
\begin{equation}\label{eq:generalziedCK}
	\mathcal{C}_{\rm gen}= -\frac{1}{\GN} \int d^d\sigma \,\sqrt{h}\,K\,.
\end{equation}
Two distinguishing features of this observable are: first, as noted above, the growth rate will remain finite even if the extremal surface is drawn to the singularity; and second, no additional scale is required to produce a dimensionless observable. Recall that typically additional factors of the AdS scale (or some other length scale) appear in the codimension-one observables, including complexity=volume. The analysis of this observable is similar to the cases described in section \ref{recap} where determining the radial profile of the extremal surface is recast as a classical mechanics problem, \eg recall eqs.~\eqref{eq:Hamiltonian} and \eqref{potent}. However, an added complication that comes from observables involving the extrinsic curvature is that the corresponding functional involves second derivatives of the profile, as discussed in appendix B of \cite{Belin:2022xmt}. Following \cite{Belin:2022xmt}, we briefly analyze the extremal surfaces with respect to this functional \eqref{eq:generalziedCK} in appendix \ref{sec:extremalK}. 

Of course, the discussion of singularities here applies to solutions of the classical Einstein equations. It would be interesting to better understand how complexity could reveal stringy or quantum corrections to the structure of the singularities \cite{Nally:2019rnw}, or possibly even how the singularities are resolved in a full solution of string theory  \cite{Khoury:2001bz,Balasubramanian:2002ry,Cornalba:2002fi,Berkooz:2002je,Simon:2002ma,Elitzur:2002rt,Horowitz:2002mw,Cornalba:2002nv,Horowitz:2003he}.

\subsection*{More on circuit complexity}
Of course, the key challenge for the complexity=anything proposal is to translate these interesting bulk observables into observables in the boundary theory,\footnote{For example, see \cite{Fidkowski:2003nf,Festuccia:2005pi} for probing the singularity inside AdS black holes using geodesics.} and thus establish a dictionary between the geometry of the black hole interior and the behaviour of boundary complexity. As a step in this direction, we can use the construction in \cite{Belin:2018fxe,Belin:2018bpg,Belin:2022xmt} to relate the CMC slice observables considered in section \ref{sec:CMC} to the symplectic form $\Omega(\delta, \delta_w)$. Here the conjugate variation $\delta_w$ is determined by the choice of gravitational observables. Small variations in the parameter $\gamma$ (or $\alpha_+$) correspond to a smooth transition between CMC slices with slightly different values of the extrinsic curvature. 
This variation can thus be interpreted as the one used to construct the gravitational symplectic $\Omega(\delta, \delta_w)$ on the semi-classical phase space. Since the bulk symplectic form naturally maps onto the boundary CFT \cite{Belin:2018fxe,Belin:2018bpg}, one naturally obtains the dual description for the variation of the gravitational observables.

More generally, we might ask what lessons for boundary theory can be drawn from our investigations of these new bulk observables. Here and in \cite{Belin:2021bga,Belin:2022xmt}, two criteria are used to identify candidates for holographic complexity: linear growth at late times and the switchback effect. However, it may still be that only a subset of our infinite family of candidates corresponds to quantum complexity in boundary theory. That is, quantum complexity can be expected to exhibit additional properties beyond linear growth and the switchback effect. One simple additional requirement is that holographic complexity must be positive for all backgrounds. This positivity constraint would actually only constrain the behaviour of the gravitational observables near the asymptotic limit, since the value of the complexity is dominated by the UV contributions. However, this constraint may still exclude certain candidates from consideration. It is an interesting question to identify other fundamental properties of quantum complexity that can serve to further constrain our infinite family of gravitational observables.

In this context, one might require that the integrand appearing for the codimension-one observables in eq.~\reef{eq:Ccodimensionone} is positive everywhere. This constraint would be motivated by imagining that there is a {\it local} map from the extremal surface in the bulk to the quantum circuit preparing the boundary state \cite{Milsted:2018yur,Milsted:2018san}. If this were the case, then just as every gate in the circuit contributes positively to the complexity, one would require that every part of the bulk surface contributes positively to the complexity. This would certainly place a strong restriction on the possible functionals that could appear in eq.~\reef{eq:Ccodimensionone}.  However, as appealing as it would be to identify the bulk surfaces and the boundary circuits, we think it unlikely that such an identification can be made by a local mapping. A fundamental obstacle seems to be the different extremization procedures in bulk and boundary. That is, in the bulk we need to maximize a certain quantity, whereas in the boundary we are looking for a minimum. This discrepancy is further highlighted in complexity=anything, since in many situations there may be more than one locally extremal surface, and one chooses the surface that maximizes eq.~\reef{eq:Ccodimensionone}.\footnote{Note that if the CV proposal were applied in Euclidean spacetimes \cite{Takayanagi:2018pml,Hernandez:2020nem}, the natural extremization procedure would be to minimize the volume, which is more akin to minimizing the circuit length in quantum complexity. A similar approach could be applied with complexity=anything, but restricting our attention to Euclidean geometries would not allow one to study black hole interiors or the time evolution of holographic complexity. See also \cite{Caputa:2017urj,Caputa:2017yrh,Boruch:2020wax,Boruch:2021hqs,Pedraza:2021mkh} for other approaches to complexity with minimization procedures.}

One interesting feature of the complexity=anything proposal is that the gravitational observables can be constructed with two independent sets of functionals. For example, in eq.~\reef{eq:O1} we have $(F_{1\pm}, G_1)$ in the observable evaluated on the extremal region and $(F_{2\pm}, G_2)$ in the extremization procedure. We reiterate that this feature was central to the construction of a family of gravitational observables with which we were able to construct a detailed picture of the black hole interior in section \ref{sec:CMC}. Perhaps there is a lesson here for quantum complexity. That is, the central question in quantum complexity is to identify the optimal (or shortest) quantum circuit for a given task. However, once the optimal circuit has been identified, it may be interesting to characterize various properties of the circuit beyond it's length. This feature of complexity=anything would have a natural counterpart in Nielsen's geometric approach to quantum complexity \cite{Nielsen:2006,nielsen2006quantum,nielsen2008}. There, the question of determining the optimal circuit is translated into finding an extremal path through a given geometry, but given the extremal path, we can easily imagine evaluating other geometric features of the same path, \ie applying other measures to characterize the optimal circuit. Of course, this raises the interesting question of what are useful measures of optimal quantum circuits.

Returning to our gravitational observables, if the above interpretation is correct, it would suggest that the quantum complexity would be related to the extremization procedure and the extremal value of the corresponding quantity, \eg the observable \reef{eq:O1} evaluated with $(F_{1\pm}, G_1)=(F_{2\pm}, G_2)$. Other choices of $(F_{1\pm}, G_1)$ would be measuring other properties of the underlying optimal circuit. From this perspective, it may seem surprising that these new measures characterizing the optimal circuit exhibit complexity-like features such as late-time linear growth.

Now it is obvious that the ability to explore the vicinity of the singularity region depends only on the position of the extremal surfaces, and thus only on the choice of functionals used in the extremization process. We emphasize that by varying the position of the surface, or by varying these functionals, we are exploring a {\it single} boundary state in different ways. That is, we are choosing different measures that may make different geometric features within the WDW patch of a single boundary state more manifest. From a bulk perspective this is perhaps not so surprising, \ie given the appropriate initial data on any Cauchy surface anchored to a fixed time slice in the boundary, we can evolve the bulk equations of motion to determine the full spacetime geometry throughout the WDW patch. However, we find it intriguing that complexity=anything suggests that quantum circuits preparing the same boundary state, but optimized in different ways, reveal different geometric features of the dual bulk geometry.

Another feature that deserves some discussion from a boundary perspective is the unusual behaviour found in section \ref{sec:end} when various parameters defining the gravitational observables became `too large'. These behaviours were illustrated with the $C^2$ observable in eq.~\reef{eq:generalziedC2}, for which locally extremal surfaces survived only within a relatively narrow range of the parameter $\lambda$. For positive $\lambda$, the maximal surface is pushed to the boundary of phase space, \ie to the upper boundary of the WDW patch, and $\mathcal{C}_{\rm gen}$ diverges -- unless regulated as described in section \ref{regular1}. A speculative suggestion would be that in this parameter regime, a certain class of gates essential for state preparation are eliminated from the underlying quantum circuits. The regularization procedure could then allow these or related gates, but with a very large penalty. This would allow the desired boundary state to be prepared, but at a large complexity. Further, the complexity would diverge again, but in a controlled way as the regulator is removed. From the perspective of the above discussion, it is interesting that these circuits with `divergent' complexity can still be characterized by other measures, \eg evaluating the observable in eq.~\reef{eq:generalziedCK} on these surfaces would yield a finite growth rate even after the regulator is removed.

The other interesting behaviour found in section \ref{sec:end} arose when $\lambda$ became too negative. In this situation, there were no contributions from behind the horizon because $a(r)$ becomes negative there. While again we can only offer speculative remarks, this behaviour could be interpreted as indicating that in this parameter regime, new `complex' gates are introduced which trivialize the infrared structure of the boundary state. To better explain this remark, we make an analogy with the description of a thermofield double (TFD) state of a two-dimensional CFT, which results from tensor network renormalization \cite{evenbly2015tensor,evenbly2015tensor2}. The tensor network includes two standard MERA circuits \cite{Vidal:2007hda,vidal2009entanglement,Haegeman:2011uy,Nozaki:2012zj} which establish the correct short-range correlations within each component of the TFD state. There is also a central set of tensors\footnote{Note that these tensors are not unitary gates.} which connect these two circuits and provide the TFD state with the correct thermal spectrum. Including these central tensors among the allowed gates in the underlying quantum complexity model would be an example of trivializing the infrared structure of the boundary state. In the AdS Schwarzschild black hole studied in this paper, $C^2$ observables with large negative $\lambda$ prevent late-time linear growth. An interesting way to explore these ideas would be to see if the unusual behaviour of the bulk observable extends to all backgrounds, \eg to see if there are black hole solutions where $C^2$ observables with large negative $\lambda$ could grow at late times.

%%%%%%%%%%%%%%%%%%%%%%%%%%%%%%%%%%%%%%%%%%%%%%
%%%%%%%%%%%%%%%%%%%%%%%%%%%%%%%%%%%%%%%%%%%%%%

%\newpage

\begin{acknowledgments}
We are happy to thank Alex Belin, Aidan Herderschee, Ted Jacobson, Finn Larsen, Gabor Sarosi and Antony Speranza for fruitful discussions and useful comments.  Research at Perimeter Institute is supported in part by the Government of Canada through the Department of Innovation, Science and Economic Development Canada and by the Province of Ontario through the Ministry of Colleges and Universities. RCM is supported in part by a Discovery Grant from the Natural Sciences and Engineering Research Council of Canada, and by funding from the BMO Financial Group.  RCM and SMR are supported by the Simons Foundation through the ``It from Qubit'' collaboration. SMR is also supported by MEXT-JSPS Grant-in-Aid for Transformative Research Areas (A) ``Extreme Universe'', No. 21H05187 and by JSPS KAKENHI Research Activity Start-up Grant Number JP22K20370.
\end{acknowledgments}

%%%%%%%%%%%%%%%%%%%%%%%%%%
%%%%%%%%%%%%%%%%%%%%%%%%%%

\appendix
%%%%%%%%%%%%%%%%%%%	%%%%%%%%%%%%%%%%%%%		
\section{Finite GHY boundary term on the singularity}\label{sec:nearsingularity}
%%%%%%%%%%%%%%%%%%%	%%%%%%%%%%%%%%%%%%%	
Due to the nature of the cosmological/spacelike singularity, many scalar functionals constructed from the bulk Riemannian tensors, such as the Weyl square term $C^2$ are divergent at the singularity. Nevertheless, one can check that the Gibbons-Hawking-York boundary term $K\sqrt{h}$ remains finite for a wide range of bulk spacetimes containing the spacelike singularity. This is due to the fact that while the volume measure $\sqrt{h}$ on the singularity vanishes, the trace of the extrinsic curvature of the singularity diverges, for instance in the null limit of the constant mean curvature slice. It is worth noting that the integrand appearing in the general codimension-one functionals \eqref{eq:obsdef} or codimension-zero functionals \eqref{eq:O1} may differ from those used to determine the extremal surface. As a consequence, the GHY boundary term provides a finite measure for the holographic complexity even when the extremal surface approaches the singularity. This appendix aims to prove the finiteness of the GHY boundary term on the spacelike singularity, under the assumption that the energy-momentum stress tensor is not rapidly divergent.

Supposing we are interested in the asymptotic geometries of the cosmological singularity, we can define the corresponding Gauss normal coordinates as follows
\begin{equation}
	ds^2 = - d\tau^2 + h_{ij}(\tau,x^i) dx^i dx^j\,.
\end{equation}
Here, the singularity is situated at $\tau=0$, and the normal vector of the spacelike hypersurface at this point is given by $n^\mu = (1, \vec{0} ) $. The advantage of these Gauss normal coordinates lies in their simplicity, allowing us to directly calculate the extrinsic curvature of the spacelike singularity, which is given by
\begin{equation}\label{eq:KijGauss}
	\begin{split}
		K_{ij} &= \frac{1}{2} \partial_\tau h_{ij} \,, \qquad K = \frac{1}{\sqrt{h}} \partial_\tau \sqrt{h} = \partial_\tau \(  \ln \sqrt{h} \)\,,
	\end{split}
\end{equation} 
For the latter purpose, it is worth noting that the Gauss, Codazzi, and Ricci equations in Gauss normal coordinates take simplified forms, \viz 
\begin{equation}
	\begin{split}
		\tensor{\mR}{_i_j_k_l} &= \bar{R}_{ijkl} - \epsilon \(  K_{ik}K_{jl} - K_{jk}K_{il}  \) \,, \\ 
		\tensor{\mR}{_i_j_k_\tau} &= D_i K_{jk} - D_j K_{ik} \,, \\
		\tensor{\mR}{_i_\tau_j_\tau} &= - \partial_\tau K_{ij} + \tensor{K}{_i^k} \tensor{K}{_k_j}\,, \\
	\end{split}
\end{equation}
where $ \bar{R}_{ijkl}$ denotes the intrinsic Riemann curvature tensor on the hypersurface located at $\tau=0$ and the covariant derivative $D_i$ is associated with the induced metric $h_{ij}$.

To analyze the asymptotic geometry in the vicinity of the spacelike singularity, we consider the following asymptotic expansion 
\begin{equation}
	\lim\limits_{\tau \to 0} \sqrt{h} \approx F(x^i)+G(x^i)\tau^\Delta + \mathcal{O}(\tau^{\Delta+1}) \,, 
\end{equation}
when a spacelike hypersurface approaches the spacelike singularity at $\tau=0$. Since the size of the spacelike singularity vanishes, it follows that  $F(x^i)=0$ and $\Delta >0$ as the constraints. The trace of the extrinsic curvature near the singularity is dominated by 
\begin{equation}
	\lim\limits_{\tau \to 0}  K \sqrt{h} \equiv  \lim\limits_{\tau \to 0}    \partial_\tau \sqrt{h}   \approx  \Delta  G(x^i) \tau^{\Delta-1} + \mathcal{O}(\tau^{\Delta}) \,. 
\end{equation}
Therefore, the finiteness of the GHY term at the spacelike singularity, \ie 
\begin{equation}
	\lim\limits_{\tau \to 0}  K \sqrt{h}  = \text{Finite Constant} = G(x^i) \ne 0 \,,
\end{equation}
is equivalent to the requirement
\begin{equation}
	\Delta = 1 \,,
\end{equation}
which determines the rate at which the volume shrinks to zero near the singularity. 

Of course, we would like to interpret this property from a more physical viewpoint by relating it to the constraint of the matter stress tensor. By taking the normal derivative of $K \sqrt{h}$, we obtain
\begin{equation}\label{eq:Kh}
	\begin{split}
	n^\mu \nabla_\mu \( K\sqrt{h} \) &= \partial_\tau \(  K \sqrt{h} \) = \partial_\tau \partial_\tau \sqrt{h}   \\
	& = \frac{\sqrt{h}}{2} \(  \frac{1}{2} ( h^{ij}\partial_\tau h_{ij})^2  + \partial_\tau h_{ij}  \partial_\tau h^{ij}   +h^{ij} \partial_\tau \partial_\tau h_{ij} \)   \,.  \\
	\end{split}
\end{equation}
From the series expansion of the volume measure near the singularity, it is straightforward to get
\begin{equation}\label{eq:Kexpansipn}
	\lim\limits_{\tau \to 0}   n^\mu \nabla_\mu \( K\sqrt{h} \)  \approx G(x^i) \Delta \( \Delta -1 \) \tau^{\Delta-2} + \mathcal{O}(\tau^{\Delta-1})  \,. 
\end{equation}	
On the other hand, contracting the Gauss-Codazzi equation gives rise to 
\begin{equation}
	\begin{split}
		\mR_{ij} &= \bar{R}_{ij} + \partial_\tau K_{ij} + KK_{ij} - 2 \tensor{K}{_i_k}\tensor{K}{^k_j} \,, \\
		\mR_{ij} h^{ij} &= \bar{R} +  h^{ij} \partial_\tau K_{ij} +K^2 - 2 K_{ij}K^{ij} \,.
	\end{split}
\end{equation}
Substituting the expressions of the extrinsic curvature in Gauss normal coordinates, \ie eq.~\eqref{eq:KijGauss}, we can find 
\begin{equation}\label{eq:RijR}
	\mR_{ij} h^{ij} -  \bar{R} = \frac{1}{2} h^{ij} \partial_\tau\partial_\tau h_{ij}+ \frac{1}{4} ( h^{ij}\partial_\tau h_{ij})^2  + \frac{1}{2}  \partial_\tau h_{ij}  \partial_\tau h^{ij}  \,.
\end{equation}
Combining the two similar equations in eqs.~\eqref{eq:Kh}, \eqref{eq:RijR} and the equivalence in eq.~\eqref{eq:Kh}, we arrive at the following identifications: 
\begin{equation}\label{eq:identification}
	\begin{split}
		n^\mu \nabla_\mu \( K\sqrt{h} \) =\partial_\tau \partial_\tau \sqrt{h}  &= \sqrt{h} \(   \mathcal{R}_{ij} h^{ij}  - \bar{R} \) = \sqrt{h} \(   \mathcal{R} + \mathcal{R}_{\tau\tau}  - \bar{R} \) \,.\\
	\end{split}
\end{equation}
Taking the limit near the singularity, one can expect the following series of expansion 
\begin{equation}
	\lim\limits_{\tau \to 0}  \(   \mathcal{R}_{ij} h^{ij}    - \bar{R} \) \approx   \lim\limits_{\tau \to 0}  \(   \mathcal{R} + \mathcal{R}_{\tau\tau} \)  \approx \frac{N(x^i)}{\tau^2} + \cdots + \frac{N'(x^i)}{\tau} + \mathcal{O}(\tau^0) \,, 
\end{equation}	
where the intrinsic curvature of the spacelike singularity at $\tau=0$ is negligible to the leading order. We note that the leading term is always at the order of $\mathcal{O}(\tau^{-2})$ because the Riemannian tensors contain at most two derivatives of the metric components \footnote{One can also confirm this by using the equality in eq.~\eqref{eq:identification} and the expansion in eq.~\eqref{eq:Kexpansipn}.}. 
From the equality derived in eq.~\eqref{eq:identification}, one can fix the coefficient of the leading term, \viz 
\begin{equation}
	N (x^i)= \Delta \( \Delta -1 \).  
\end{equation}
by using the series expansion eq.~\eqref{eq:Kexpansipn}. We remark that this equivalence is a geometric result without taking any other assumptions. 

On the other hand, Einstein equation of $(d+1)$-dimensional bulk spacetime tells us that the matter stress tensor in the vicinity of the singularity is expressed as:
\begin{equation}\label{eq:stresstensor}
  \lim\limits_{\tau \to 0}  \(   T_{\tau\tau} - \frac{T_\mu^\mu}{d-1} \) \approx \lim\limits_{\tau \to 0}  \(   \mathcal{R} + \mathcal{R}_{\tau\tau} \)  \approx \frac{N}{\tau^2} + \cdots + \frac{N'(x^i)}{\tau} + \mathcal{O}(\tau^0) \,. 
\end{equation}
We assume that the energy-momentum stress tensor\footnote{Of course, we only need to constrain the combination  $T_{\tau\tau} - \frac{T_\mu^\mu}{d-1} $ here. Taking the perfect fluid with $T_{\mu\nu}= (\rho + p)n_\mu n_{\nu} + (p +\rho) g_{\mu\nu} $ as an example, we have $T_{\tau\tau} - \frac{T_\mu^\mu}{d-1} = \frac{d}{d-1} \(  \rho - p \)$.} is not rapidly divergent near the singularity, \ie its potential leading divergence should vanish with 
\begin{equation}
	N=0  \,. 
\end{equation}
To put it another way, it simply means that the divergence of stress tensor with approaching the singularity is constrained by 
\begin{equation}\label{eq:mattercondition}
	 \lim\limits_{\tau \to 0}  \(   T_{\tau\tau} - \frac{T_\mu^\mu}{d-1} \)  \ll \frac{1}{\tau^2} \,. 
\end{equation}
Adhering to this condition, we can immediately conclude $\Delta=1$, which implies that the GHY boundary term $K\sqrt{h}$ on the spacelike singularity is finite as we advertised before. The condition eq.~\eqref{eq:mattercondition} we imposed on the energy-momentum stress tensor is also part of the original BKL conjecture that the matter could be neglected asymptotically in the neighborhood. 

To further illustrate this condition, let us consider the Kasner metric
\begin{equation}
		ds^2 =  -d\tau^2 + \sum_i^d \tau^{2p_i} dx_i^2   \,. 
\end{equation}
as a practice. It is easy to show 
\begin{equation}
	  \mathcal{R}_{ij} h^{ij}  - \bar{R} =   \mathcal{R}_{ij} h^{ij}   = \(  p_1 + p_2 +\cdots + 
	  p_d  \)\(  p_1 + p_2 +\cdots+ p_d  -1 \)\frac{1}{\tau^2} \,. 
\end{equation}
Our assumption $N=0$ appearing in eq.~\eqref{eq:stresstensor} is equivalent to the constraint for Kasner geometry, \ie 
\begin{equation}
	\sum_i^d p_i = 1 \,. 
\end{equation}
Of course, one can check that this condition guarantees the finiteness of the GHY boundary term on the Kasner singularity: 
\begin{equation}
	\lim\limits_{\tau \to 0} K \sqrt{h} = \frac{p_1 + p_2 +\cdots + 
		p_d  }{\tau} \tau^{p_1 + p_2 +\cdots + 
		p_d} =1 \,. 
	\end{equation}
%%%%%%%%%%%%%%%%%%%%%%%%%%%%	
%%%%%%%%%%%%%%%%%%%%%%%%%%%%	
%%%%%%%%%%%%%%%%%%%%%%%%%%%%	
%%%%%%%%%%%%%%%%%%%	%%%%%%%%%%%%%%%%%%%		
\section{Extremal Surfaces for the Extrinsic Curvature}\label{sec:extremalK}
%%%%%%%%%%%%%%%%%%%	%%%%%%%%%%%%%%%%%%%	
In this appendix, we investigate a simple codimension-one observable defined by the extrinsic curvature of a hypersurface, which is expressed as 
\begin{equation}\label{eq:generalziedCK02}
	\mathcal{C}_{\rm gen}= -\frac{1}{\GN} \int d^d\sigma \,\sqrt{h}\,K\,.
\end{equation}
where $K$ is the trace of the extrinsic curvature of the hypersurface. Given the bulk spacetime as the general AdS black hole defined by eq.~\eqref{eq:AdSBH}, the trace of the extrinsic curvature of a hypersurface parametrized by $(v(\sigma), r(\sigma))$ reads 
\begin{equation}\label{eq:defineK}
	\begin{split}
		K &=    \frac{ 4(d-1) \dot{v}\dot{r}^2 - \left( 2(d-1)f(r) +r f'(r)  \right) \(3\dot{r}- f(r) \dot{v}\)\dot{v}^2      - 2 r\( \dot{r}\ddot{v}-\dot{v}\ddot{r}\) }{2 r (2 \dot{v} \dot{r}-f(r) \dot{v}^2)^{3/2}  }  \,. 
	\end{split}
\end{equation}
For example, it reduces 
\begin{equation}
K \big|_{r=r_0<r_h}=\frac{2(d-1) f(r_0) +r_0f'(r)}{ 2r_0\sqrt{-f(r_0) }} \,.
\end{equation}
for the constant radius slice inside the horizon. 
In comparison with the functionals analyzed in section \ref{codone}, the novel characteristic associated with the above functional \eqref{eq:generalziedCK02} is the appearance of the second-order derivative terms stemming from the extrinsic curvature. This phenomenon has been explored in detail in appendix B of \cite{Belin:2022xmt}. Consequently, the conjugate momentum is altered as 
\begin{equation}\label{eq:PvK}
	\begin{split}
		P_v &\equiv \frac{\partial \mathcal{L}_{\rm gen}}{\partial \dot{v}} - \frac{d}{d\sigma} \(  \frac{\partial \mL_{\rm gen}}{\partial \ddot{v}}   \)= - \(\frac{r}{L}\)^{d-2}  \(  (d-1) f(r)+ \frac{rf'(r)}{2}\)   + (d-1)\(\frac{L}{r}\)^{d} \dot{r}^2 \,, 
	\end{split}
\end{equation} 
where we have used the gauge condition \reef{eq:gauge01} as before. The time derivative of the observable \eqref{eq:generalziedCK02} with respect to the boundary time $\tau$ is still controlled by the conserved momentum as 
\begin{equation}\label{eq:dVdt03}
	\frac{d \mathcal{C}_{\rm gen}}{d\tau} =\frac{\Omega_{k, d-1} L^{d-2}}{\GN}\,P_v(\tau)\,. 
\end{equation}
To illustrate the corresponding extremal surfaces, we consider the planar black hole where $f(r)$ is given in eq.~\reef{blacken} with $k=0$. The corresponding extremization equation can be cast as: 
\begin{equation}
	\dot{r}^2\equiv - \mathcal{U}(P_v, r) =   \frac{w}{2(d-1)}  \(  d \(\frac{r_h}{L} \)^d  (2w-1)  + 2 P_v   \)  \(\frac{r_h}{L} \)^d \,,
\end{equation}
where the horizon radius and the dimensionless radial coordinate are given by $r_h^d=L^2 \omega^{d-2}$ and $w= (r/r_h)^d$, respectively. We can deduce from the radial equation that there exist three types of extremal surfaces. The turning point or minimal radius is given by 
\begin{equation}
w_{\rm min}=  \(\frac{r_{\rm min}}{r_h} \)^d =  \frac{1}{2} - \frac{P_v}{d} \( \frac{L}{r_h} \)^d \,. 
\end{equation}
Of particular interest are the extremal surfaces that are capable of crossing the horizon and connecting the left and right boundaries. This is achieved by selecting the conserved momentum from the following range:
\begin{equation}
	P_v \in \[  - \frac{d}{2}  \( \frac{r_h}{L}\)^d,   \frac{d}{2}  \( \frac{r_h}{L}\)^d\] \,. 
\end{equation}
In particular, the maximum momentum corresponds to the vanishing extremum of the effective potential $\mathcal{U}(P_v, r)$, indicating that the final slice anchored at $\tau \to \infty$ coincides with the spacelike singularity at $w=0$ or $r=0$. Further the late-time growth rate is given by $\frac{d\mathcal{C}^+}{d\tau} \simeq\frac{8\pi M\,d}{(d-1)}$, just as in eq.~\reef{Kobs}. For $P_v \le - \frac{d}{2} \(\frac{r_h}{L}\)^d$, the extremal surfaces are located outside the horizon. Conversely, if $P_v > \frac{d}{2} \(\frac{r_h}{L}\)^d$, the extremal surfaces originating at the left/right boundary would collide with the singularity.

%%%%%%%%%%%%%%%%%%%%%%%%%%
%%%%%%%%%%%%%%%%%%%%%%%%%%
\bibliographystyle{jhep}
\bibliography{bibliography}
%%%%%%%%%%%%%%%%%%%%%%%%%%
%%%%%%%%%%%%%%%%%%%%%%%%%%

\end{document}